\documentclass[12pt]{article}
\usepackage{amssymb}
\usepackage{amsfonts}
\usepackage{amsmath}
\usepackage{latexsym}
\usepackage{xcolor}
\usepackage{color}
\usepackage{graphics}
\usepackage{graphicx}
\usepackage[T1]{fontenc}
\usepackage[utf8]{inputenc}
\usepackage{subfigure}
\usepackage{float}
\begin{document}
\title{{
Cosmological "constant" in a universe born \\ in the metastable false vacuum state
}}
\author{K. Urbanowski\footnote{e--mail:  K.Urbanowski@if.uz.zgora.pl, $\;$ k.a.urbanowski@gmail.com} \\
University of Zielona G\'{o}ra, Institute of Physics, \\
ul. Prof. Z. Szafrana 4a,  65--516 Zielona G\'{o}ra, Poland.}
\maketitle

\begin{abstract}
The cosmological constant $\Lambda$ is a measure of the energy density of the vacuum. Therefore properties of the energy of the system in the metastable vacuum state reflect
properties of $\Lambda = \Lambda(t)$.  We analyze
properties of the energy, $E(t)$,  of a general quantum system in the metastable state in various phases of the decay process: In the exponential phase, in the transition phase
between the exponential decay and the later phase, where decay law as a function of time $t$ is in the form of powers of $1/t$, and also in this last phase.
We found that this energy having an approximate  value resulting from the Weisskopf--Wigner theory in  the exponential decay phase is reduced very fast in the transition phase to its
asymptotic value $E(t) \simeq E_{min} + \alpha_{2}/t^{2}+\ldots$ in the late last phase of the decay process. (Here $E_{min}$ is the minimal energy of the system). This quantum
mechanism  reduces the energy of the system in the unstable state by a dozen or even several dozen orders or more.
We show that if to assume that a universe was born in metastable false vacuum state then according to this quantum mechanism
 the cosmological constant $\Lambda$ can have a very great value  resulting from the quantum field theory calculations in the early universe in the inflationary era, $\Lambda \simeq
 \Lambda_{qft}$,
 and then it can later be quickly reduced to the very, very small values.
 \end{abstract}
\noindent

\section{Introduction }

Many physical processes including some cosmological processes are quantum decay processes.
Attempts to solve the problem of a description of  the evolution in time and decay of quantum unstable (or metastable) states were made
practically since the birth of the Quantum Theory.
Difficulties with this problem are caused by the fact that unstable states are not eigenvectors of the self--adjoint Hamiltonian $\mathfrak{H}$ governing the time evolution in the
system containing such states. The problem is important because one can meet unstable  (or metastable) states in many quantum processes: Starting from the spontaneous emission of
electromagnetic radiation by excited quantum levels of molecules or atoms \cite{WW},  through the radioactive decay of radioactive elements (e. g., $\alpha$--decay \cite{Gamov}),
decays of almost all known elementary particles,  to the problem of the false vacuum decay, which is a quantum process \cite{Coleman:1977py,Callan:1977pt}.
Therefore if one wants to search for properties of the universe born in the false vacuum state, one must know how to describe the quantum decay process of such a state.

In fact, the problem of describing the decay process appeared as early as in the pre-quantum theory era, when it was necessary to quantify the changes in  time of a number of decaying
radioactive elements.  The radioactive decay law formulated by Rutherford and Sody in the nineteenth century
\cite{rutheford,rutheford1,rutheford2}
allowed to determine the number $N (t)$ of  atoms   of the radioactive element at the instant $t$  knowing the initial number  $N_{0} = N(0)$ of them  at  initial instant of time
{$t_{0}^{init} =0$} and had the exponential form: $N(t) =N_{0}\,\exp\,[-\lambda t]$, where
where {$\lambda > 0$} is a constant.
Since then, the belief that the decay law has the exponential form has become common.
This conviction was upheld by Wesisskopf--Wigner theory of spontaneous emission \cite{WW}: They found that to a good approximation the quantum mechanical non--decay probability of the
exited levels is a decreasing function of time having exponential form.
Further studies of the quantum decay process showed that basic principles of the quantum theory  does not allow it to be described by an exponential decay law at very late times
\cite{Khalfin,Fonda} and at initial stage of the decay process (see e. g.  \cite{Fonda} and references therein). Theoretical analysis shows that at late times the survival probability
(i. e. the decay law) should tends to zero as
{$t \to \infty$} much more slowly than any exponential function of time and that as function of time it has the inverse power--like form
at this  regime of time \cite{Khalfin,Fonda}.
All these results caused that there is rather widespread belief that a universal feature of the quantum decay process is the presence of three
time regimes of the decay process:
the early time (initial), exponential (or "canonical"), and late time having inverse--power law form \cite{Peshkin}.
This  belief is reinforced by a numerous presentations in the literature of decay curves obtained for quantum models of unstable systems.

The theoretical studies of unstable states mentioned resulted in discovery some new quantum effects, such as the Quantum Zeno (and anti--Zeno) effects \cite{misra,bh,bh1,kk,pf}
resulting from early time properties of the time evolution of the unstable state, and
which were confirmed experimentally \cite{WMI}, or a reduction of the energy of the system in the unstable state at late times
\cite{Urbanowski:2006mw,Urbanowski:2008kra,Urbanowski:2009lpe,Urbanowski:2011zz,Urbanowski:2016pks}, which is connected with  the asymptotic late time  behavior of the survival
probability.


As it was already mentioned,
the theory of quantum decay processes has found its applications in cosmology: e.g. in the studies of cosmological models in which there is a metastable false vacuum
and, consequently, the decaying dark energy. Coleman et al. in  seminal papers \cite{Coleman:1977py,Callan:1977pt}
discussed the instability of a physical system, which is not at an absolute energy minimum, and which is separated from the absolute minimum by an effective potential barrier. They
showed that if the early Universe is too cold to activate the energy transition to the minimum energy state then a quantum decay, from the false vacuum to the true vacuum, is still
possible through a barrier penetration via the macroscopic quantum tunneling. In other words they showed that quantum decay processes can play an important role in the early Universe.
What is more, it appears that asymptotic late time properties of of quantum decay processes can be responsible for some cosmological effects.
This idea was formulated by Krauss and Dent \cite{Krauss:2007rx,Krauss:2008pt}. They analyzing
a false vacuum decay pointed out that in eternal inflation,
many false vacuum regions can survive up to the times much later than times when the exponential decay law holds.
Krauss and Dent gave a simple explanation of this effect:
It may occur even though regions of false vacua by assumption
should  decay exponentially, gravitational effects
force space in a region that has not decayed yet
to grow exponentially fast. In general,
the space grows exponentially fast only
in the inflationary phase of the evolution of the Universe (see, e.g. \cite{Cheng,Weinberg-cos}). Therefore a realization of  Krauss and Dent's hypothesis in our Universe
is possible only if the lifetime of the false vacuum is much shorter than the duration of the inflationary epoch because only then the quantum decay of the false vacuum takes place in
its canonical regime.
The mentioned Krauss and Dent's idea was used in \cite{Urbanowski:2011zz,Urbanowski:2016pks} to show that when analyzing cosmological processes not only late time properties of the
survival probability of  decaying false vacua should be considered but  also, what seems even more important,
the late time properties of the energy in the false vacuum state of the system. In cosmology, when we study the decay of a false vacuum, the Universe is the quantum system under
consideration.

The late time effect considered by Krauss and Dent \cite{Krauss:2007rx} is
impossible within the standard approach of calculations of decay rate $\Gamma$ for decaying vacuum state (see e.g., \cite{Coleman:1977py,Callan:1977pt} and many other papers).
Calculations performed within this standard approach cannot lead to a correct description of the evolution of the Universe with false vacuum in all cases when the lifetime of the
false vacuum state is such short that its survival probability exhibits an inverse power-law behavior at times which are of the order of the age Universe or  shorter.
 This conclusion is valid not only when the dark energy density and its late time properties are related to the transition of the Universe from the false vacuum state to the true
 vacuum but also when the dark energy is formed by unstable "dark particles". In both cases the decay of the dark energy density is the quantum decay process and only the formalism
 based on the Fock--Krylov theory of quantum unstable states and used by Krauss and Dent \cite{Krauss:2007rx} is able to describe correctly such a situation.

In the analysis performed in this paper we will assume the model of the dark anergy close to that considered by Landim and Abdalla,
in which the observed vacuum energy is the value of the scalar potential at the false vacuum \cite{Landim:2016isc}. (Similar idea was used in many papers --- see eg.
\cite{Stojk,Lim}). In  other words, we will assume that
the current stage of accelerated expansion of the universe will be described by a canonical scalar field $\Phi$  such that its potential, $V(\Phi)$, has a local and true minimums. So,
the field at the false vacuum will represent the darkenergy. In such a situation,
the quantum state of the system in the local minimum is described by a state vector corresponding to the false vacuum state whereas the quantum state of the system in the true minimum
corresponds to the state of the lowest energy of the system and it is a true vacuum.
This means that the density of the energy of the system in the false vacuum state, $\rho_{vac}^{\,\text{F}}$, will be  identified with the density of the dark energy,
$\rho_{vac}^{\,\text{F}} \equiv \rho_{de}$, (or, equivalently, as the cosmological term $\Lambda$) in the Einstein equations \cite{Cheng,Weinberg-cos}. Implications of the assumption
that, $\rho_{vac}^{\,\text{F}} = \rho_{de} = \rho_{de}(t)$, behaves as the asymptotically late form of the energy of the system in the  false vacuum state, i.e. cosmologies with
decaying dark energy were studied, e. g.,  in \cite{epjc-2017b,Szydlowski:2017wlv,Szydlowski:2015fya,jcap-2020}.
In these studies, the problem of the possible reaction of the system to energy changes during the transition from the time epoch of exponential decay to the
epoch with the decay law of the form of powers of $1/t$ was not analyzed.
In particular, it was not analyzed how fast the energy tends to its asymptotic form in the era in which the decay law is proportional to the powers of $1/t$.
The aim of this paper is to investigate this problem
and also to analyze the possible influence of this effect on the currently observed properties of the system (i.e., in the case under consideration, the Universe):
Here we show that there exists a mechanism that reduces the energy of the system in the unstable state by a dozen or even several dozen orders or more, which can help to explain the
cosmological constant problem.

The paper has the following structure. Section~2 contains preliminaries: A brief introduction in the Fock--Krylov approach to the description of quantum unstable states, a brief
derivation of the effective Hamiltonian governing the time evolution of the unstable state, and short discussion of properties of the instantaneous energy of the system in the
unstable state as well as of the instantaneous decay rate is presented here for readers convenience.
Necessary model calculations and numerical results in a graphical form are presented in Section~3. Section~4 contains a discussion of possible cosmological implications results
presented in previous sections. Section~5 contains final remarks .

\section{Preliminaries}

Understanding the basic features of an unstable system requires isolating this system from the influence of the environment on the decay process, including the possible distortion of
these features by repeatedly interactions, at different times, with measuring instruments.  These conditions are met in the case of quantum decay processes  occurring in a vacuum.
Therefore, further considerations will only cover decays that occur in a vacuum.
The standard approach to study properties of quantum unstable systems decaying in the vacuum and evolving in time is to analyze their decay law (survival probability) ${\cal P}(t)$,
which describes the probability od finding the system at the instant of time $t$ in the  metastable  state $|\phi\rangle \in {\cal H}$ prepared at the initial instant $t_{0}^{init} <
t$:
\begin{equation}
{\cal P}(t) = |{\cal A}(t)|^{2}, \label{P(t)}
\end{equation}
where
\begin{equation}
{\cal A}(t) = \langle \phi| \phi(t)\rangle
\label{amp}
\end{equation}
is the survival amplitude
and $| \phi (t)\rangle $ is the solution of the Schr\"{o}dinger equation
\begin{equation}
i \hbar \frac{\partial}{\partial t} |\phi(t)\rangle = \mathfrak{H}
|\phi (t)\rangle. \label{Sch}
\end{equation}
  Here $\mathfrak{H}$ denotes the complete (full), self-adjoint Hamiltonian of the system acting in the Hilbert space ${\cal H}$ of states of
  this system
  $|\phi\rangle, |\phi (t) \rangle \in {\cal H}$, $\langle \phi|\phi\rangle = \langle \phi(t)|\phi (t) \rangle = 1$
The initial condition for Eq.~(\ref{Sch}) in the case considered is usually assumed to be
\begin{equation}
| \phi (t = t_{0}^{init} \equiv 0) \rangle \stackrel{\rm def}{=}
| \phi\rangle, \quad \text{or equivalently}, \quad {\cal A}(0) = 1. \label{A(0)}
\end{equation}
Using the basis in ${\cal H}$ build from normalized eigenvectors $|E\rangle,\,\ E\in \sigma_{c}(\mathfrak{H}) = [E_{{min}}, {\infty})$ (where $\sigma_{c}(\mathfrak{H})$ is the
continuous part of the spectrum of $\mathfrak{H} $) of $\mathfrak{H}$ and using the expansion of $|\phi\rangle$ in this basis one can express the amplitude ${\cal A}(t)$ as the
following Fourier integral
\begin{equation}
{\cal A}(t) \equiv {\cal A}(t - t_{0}^{init}) = \int_{E_{{min}}}^{\infty} \omega(E)\,
e^{\textstyle{-\,\frac{i}{\hbar}\,E\,(t - t_{0}^{init})}}\,d{E},
\label{a-spec}
\end{equation}
where $\omega(E) = \omega(E)^{\ast}$ and $\omega(E) > 0$ is the probability to find the energy of the system in the state $|\phi\rangle$ between $E$ and $E\,+\,dE$ and $E_{{min}}$ is
the minimal energy of the system. The last relation (\ref{a-spec}) means that the survival amplitude ${\cal A}(t)$ is a Fourier transform of an absolute integrable function $\omega
(E)$. If we apply the Riemann-Lebesgue lemma to the integral (\ref{a-spec}) then one concludes that there must be ${\cal A}(t) \to 0$ as $t \to \infty$. This property and the relation
(\ref{a-spec}) are an essence of the Fock--Krylov theory of unstable states \cite{Krylov:1947tmi,Fock:1978fqm}.

So, within this approach the amplitude ${\cal A}(t)$, and thus the decay law ${\cal P}(t)$ of the metastable state $|\phi\rangle$, are determined completely by the density of the
energy distribution $\omega(E)$ for the system in this state \cite{Krylov:1947tmi,Fock:1978fqm} (see also \cite{Fonda,Kelkar:2010qn}, and so on. (This approach is also applicable to
models in quantum field theory~\cite{Giacosa:2011xa,Giacosa:2018dzm}).

In \cite{Khalfin} assuming that the spectrum of $\mathfrak{H}$ must be bounded
from below and using the Paley--Wiener
Theorem \cite{Paley} it was proved that in the case of unstable
states there must be
\begin{equation}
|{\cal A}(t)| \; \geq \; A\,e^{\textstyle - b \,t^{q}}, \label{|a(t)|-as}
\end{equation}
for $|t| \rightarrow \infty$. Here $A > 0,\,b> 0$ and $ 0 < q < 1$.
This means that the decay law ${\cal P}(t)$ of metastable
states decaying in the vacuum, (\ref{P(t)}), can not be described by
an exponential function of time $t$ if time $t$ is suitably long, $t
\rightarrow \infty$, and that for these lengths of time ${\cal
P}(t)$ tends to zero as $t \rightarrow \infty$  more slowly
than any exponential function of $t$. The analysis of the models of
the decay processes shows that ${\cal P}(t) \simeq
e^{\textstyle{- \frac{{\it\Gamma}_{0} t}{\hbar}}}$, (where
${\it\Gamma}_{0}$ is the decay rate of the considered state $|\phi \rangle$),
to a very high accuracy  for a wide time range $t$: At {\em canonical decay times}, i.e., from $t$
suitably greater than some $T_{0} \simeq t_{0}^{init}= 0$ but $T_{0} >
t_{0}^{init} = 0$ (${\cal P}(t)$ has nonexponential power--like form
for short times $t \in (t_{0}^{init},T_{0})$ -- see, e.g. \cite{Khalfin,Fonda,Peres})
up to $t \gg \tau_{0} = \frac{\hbar}{{\it\Gamma}_{0}}$
and smaller than $t = T_{1}$, where $\tau_{0}$ is a lifetime and  $t=T_{1}$ denotes the
time $t$ for which the long time nonexponential deviations of ${\cal A}(t)$
begin to dominate (see eg., \cite{Khalfin}, \cite{Fonda},
\cite{Sluis}).  So, a notion {\em canonical decay times}   denotes such times $t$  that  $t \in (T_{0},T_{1})$.
From a more detailed  analysis it follows that  in the general
case there is time $T_{2} \gg T_{1}$ such that the decay law ${\cal P}(t)$ takes the inverse
power--like form $t^{- \lambda}$, (where $\lambda
> 0$), for suitably large $t \geq T_{2} \gg T_{1} \gg \tau_{0}$
\cite{Khalfin}, \cite{Fonda}, \cite{Sluis}, \cite{Goldberger}. This effect is in
agreement with the  general result (\ref{|a(t)|-as}).  Effects of this type
are sometimes called the "Khalfin effect" (see eg.
\cite{Arbo}).

The problem how to detect possible deviations from the exponential form of
${\cal P}(t)$ in the long time region  has been attracting attention of physicists since the
first theoretical predictions of such an effect \cite{Wessner,Norman1,Greenland}.
The tests that have been performed over many years to examine the form of the decay laws
for  $t \gg \tau_{0}$  have not indicated any deviations from the exponential form of ${\cal P}(t)$ in the
long time region. Nevertheless, conditions leading to the nonexponential behavior
of the amplitude ${\cal A}(t)$ at long times were studied theoretically \cite{seke} -- \cite{jiitoh}.
Conclusions following from these studies were applied successfully in experiment described  in \cite{Rothe},
where the experimental evidence of deviations from the exponential decay law at long times was
reported. This result gives rise to another problem which now becomes important:
if and how the long time deviations from the exponential decay law depend on the model considered
(that is, on the form of $\omega(E$), and
if (and how) these deviations  affect the  energy of the metastable state
and its decay rate in the long time  region.

Note that in fact the amplitude ${\cal A}(t) = \langle \phi|\phi (t)\rangle$ contains information about
the decay law ${\cal P}(t)$ of the state $|\phi\rangle$, that
is about the decay rate ${\it\Gamma}_{0}$ of this state, as well
as the energy of the system in this state.
This information can be extracted from ${\cal A}(t)$.
Using Schr\"{o}dinger equation (\ref{Sch}) one finds that within the problem considered
\begin{equation}
i \hbar \frac{\partial}{\partial t}\langle\phi |\phi (t) \rangle = \langle \phi|\mathfrak{H} |\phi (t)\rangle. \label{h||1}
\end{equation}
From this relation one can conclude that the amplitude $A(t)$ satisfies the following equation
\begin{equation}
i \hbar \frac{\partial {\cal A}(t)}{\partial t} = h(t)\,{\cal A}(t), \label{h||2}
\end{equation}
where
\begin{equation}
h(t) = \frac{ \langle \phi| \mathfrak{H} |\phi (t)\rangle }{a(t)} \equiv \frac{\langle \phi|\mathfrak{H} |\phi (t)\rangle}{\langle \phi|\phi (t)\rangle}, \label{h(t)-eq}
\end{equation}
or equivalently
\begin{equation}
h(t)  \equiv \frac{i\hbar}{{\cal A}(t)}\,\frac{\partial{\cal A}(t)}{\partial t}, \label{h(t)}
\end{equation}

The effective Hamiltonian $h(t)$ governs the time evolution in the subspace of unstable states ${\cal H}_{\parallel}= \mathbb{P} {\cal H}$, where $\mathbb{P} = |\phi\rangle \langle
\phi|$ (see \cite{Urbanowski:1994epq} and also \cite{Urbanowski:2006mw,Urbanowski:2008kra,Urbanowski:2009lpe} and references therein). The subspace ${\cal H} \ominus {\cal
H}_{\parallel} = {\cal H}_{\perp} \equiv \mathbb{Q} {\cal H}$ is the subspace of decay products. Here $\mathbb{Q} = \mathbb{I} - \mathbb{P}$. One meets the effective Hamiltonian
$h(t)$ when one starts from the Schr\"{o}dinger equation for the total state space ${\cal H}$ and looks for the rigorous evolution equation for a distinguished subspace of states
${\cal H}_{||} \subset {\cal H}$ \cite{Urbanowski:1994epq,Giraldi:2015ldu,Giraldi:2016zom}. In general $h(t)$ is a complex function of time and
in the case of ${\cal H}_{\parallel}$ of two or more dimensions
the effective Hamiltonian governing the time evolution in such a subspace is a non-hermitian matrix $H_{\parallel}$ or non-hermitian operator. There is
\begin{equation}
h(t) = E(t) - \frac{i}{2} {{\it\Gamma}}(t), \label{h-m+g}
\end{equation}
where
$
E(t) = \Re\,[h(t)]$, ${\it\Gamma}(t) = -2\,\Im\,[h(t)],
$
are the instantaneous energy (mass) $E(t)$ and the instantaneous decay rate, ${\it\Gamma}(t)$. (Here $\Re\,(z)$ and $\Im\,(z)$ denote the real and imaginary parts of $z$,
respectively).

The quantity ${\it\Gamma}(t) = -2\,\Im\,[h(t)]$ is interpreted as the decay rate, because it satisfies the definition of the decay rate used in quantum theory. Simply, using
(\ref{h(t)}) it is easy to check that
\begin{equation} \label{G-equiv}
\begin{split}
\frac{{\it\Gamma}(t)}{\hbar} &\stackrel{\rm def}{=}
- \frac{1}{{\cal P}(t)} \frac{\partial {\cal P}(t)}{\partial t} \\
&=
- \frac{1}{|{\cal A}(t)|^{2}}\,\frac{\partial |{\cal A}(t)|^{2}}{\partial t}
\equiv - \frac{2}{\hbar}\,\Im\,[h(t)].
\end{split}
\end{equation}

The  formula  (\ref{h(t)-eq}) for $h(t)$ can be used to  show that $h(t)$ can not be constant in time.
Indeed, if to rewrite the numerator
of the righthand side of (\ref{h(t)-eq}) as follows,
\begin{equation}
\langle\phi|\mathfrak{H}|\phi(t)\rangle \equiv \langle
\phi|\mathfrak{H}|\phi\rangle\,a(t)\,+\,\langle\phi |\mathfrak{H}|\phi(t)\rangle_{\perp}, \label{perp}
\end{equation}
where $|\phi(t)\rangle_{\perp} = \mathbb{Q}|\phi(t)\rangle$,
and $\langle\phi|\phi(t)\rangle_{\perp} = 0$,
then one can see
that there is a permanent contribution of
decay products described by $|\phi(t)\rangle_{\perp}$
to the energy of the metastable state considered.
The intensity of this contribution depends on time $t$.
This contribution into the instantaneous energy
is practically very small and constant
in time   to a very good approximation at  canonical decay times,
whereas at the transition times, when $t > T_{1}$ (but $t < T_{2}$, it
is fluctuating function of time and the amplitude
of these fluctuations may be significant. What is more
relations (\ref{h(t)-eq}) and (\ref{perp}) allow
one to proof that in the case of metastable
states $\Re\,[h(t)] \neq const$ for $t>0$. Namely,  using these relations
one obtains that
\begin{equation}
h(t) = E_{\phi} +\,
\frac{\langle\phi |\mathfrak{H}|\phi(t)\rangle_{\perp}}{{\cal A}(t)}, \label{h-perp-1}
\end{equation}
where $E_{\phi}$ is the expectation value of $\mathfrak{H}$: $E_{\phi} = \langle\phi|\mathfrak{H}|\phi\rangle$.
From this relation one can see that
$h(0) = E_{\phi}$ if the matrix elements
$\langle\phi|\mathfrak{H}|\phi\rangle$ exists. It is because \linebreak
$|\phi (t=0)\rangle_{\perp} =0$ and ${\cal A}(t=0)=1$.

Note now that from (\ref{h(t)-eq}) and (\ref{h(t)}) it follows that $h(t)$ must be a continuous function of time $t$ for $t\geq 0$.
So, if to assume {\em a contrario} that $h(t) = const.$ for all $t \geq 0$ then using (\ref{h(t)-eq}) and (\ref{perp}) one immediately infers that it is possible only if
 for all $t\geq 0$ there is $\langle \phi |\mathfrak{H}\mathbb{Q}|\phi (t)\rangle  \equiv c_{h}\langle \phi|\phi (t)\rangle $, where $c_{h} = const.$  From the  definition of
 $\mathbb{P}$ and $\mathbb{Q}$ it results that    in such a case there must be  ${\langle \phi |\mathfrak{H}\mathbb{Q}|\phi (t)\rangle|}_{t=0} =  \langle \phi
 |\mathfrak{H}\mathbb{Q}|\phi \rangle  = 0$, but at the same time there is, ${\langle \phi |\mathfrak{H}\mathbb{Q}|\phi (t)\rangle}|_{t=0}
 = c_{h} \langle \phi|\phi\rangle$ at $t=0$. So  only solution  is   $c_{h} \equiv 0$. Now because of the
 continuity of $h(t)$ the solution $c_{h} =0$ is valid also for all $t >0$. Thus the case $h(t) = const $ for all $t \geq 0$ occurs only if $\langle \phi ||\mathfrak{H}\mathbb{Q}|\phi
 (t)\rangle = 0$ for all $t \geq 0$. It is possible only if $[\mathbb{Q},\mathfrak{H}] \equiv [\mathbb{P},\mathfrak{H}] =0$ but then the vector $|\phi\rangle$ defining the projector
 $\mathbb{P} =|\phi\rangle\langle \phi|$ can not describe a metastable state. In a result there must be $h(t) \neq const$ for a metastable state $|\phi\rangle$.

Using projectors  $ \mathbb{P},  \mathbb{Q}$, Eq. (\ref{h(t)}) can be rewritten as follows (see, eg.  \cite{Urbanowski:2016pks,Urbanowski:1994epq})
\begin{equation}
h(t) \equiv E_{\phi} + v(t),
\end{equation}
and
\begin{equation}
v(t) =  \frac{\langle \phi|\mathfrak{H}  \mathbb{Q}|\phi (t)\rangle}{{\cal A}(t)} = \frac{\langle \phi|\mathfrak{H}|\phi (t)\rangle_{\perp}}{{\cal A}(t)}.
\end{equation}
From the definition of $\mathbb{P}$
it follows  that $|\phi (t=0)\rangle_{\perp} = 0$, which means that
$v(0) = 0$ and $h(0) = E_{\phi}$ and thus
\begin{equation}
E(0) =  E(t=0) = E_{\phi} =  \langle \phi |H| \phi\rangle,
\end{equation}
(if the matrix element $ \langle \phi |H| \phi\rangle$ exists), and
\begin{equation}
E(t) \simeq E_{\phi}=  \langle \phi |H| \phi\rangle,\;\;\;{\rm for}\;\;\; t \to 0.
\end{equation}

So, in a general case, at canonical decay times $t < T_{1}$,
there is (see \cite{Urbanowski:2016pks,Urbanowski:1994epq})
\begin{equation}
E(t) \simeq E_{0} \stackrel{\rm def}{=}  E_{\phi} -  \Delta_{\phi}^{(1)} \equiv E(0) - \Delta_{\phi}^{(1)}, \label{D1}
\end{equation}
where $\Delta_{\phi}^{(1)} = - \Re\,[v(t)]$ , wherein $v(t) \simeq const.$ at canonical decay times, and $| \Delta_{\phi}^{(1)}| \ll | E_{\phi}|$.

The representation of the survival amplitude ${\cal A}(t)$ as the Fourier transform (\ref{a-spec}) can be used to find the late time asymptotic form of ${\cal A}(t)$, ${\cal P}(t)$
and the instantaneous energy $E(t)$ and decay rate ${\it\Gamma}(t)$ (see \cite{Urbanowski:2006mw,Urbanowski:2008kra}). There is,
\begin{equation}
E(t) \underset{t \to \, \infty}{\thicksim} \stackrel{\rm def}{=} E_{lt}(t) = E_{min} + \frac{\alpha_{2}}{t^{2}} + \frac{\alpha_{4}}{t^{4}} + \ldots, \label{E(t)-as}
\end{equation}
and
\begin{equation}
{\it\Gamma} (t) \underset{t \to \, \infty}{\thicksim} \stackrel{\rm def}{=} {\it\Gamma}_{lt}(t) =  \frac{\alpha_{1}}{t} + \frac{\alpha_{3}}{t^{3}} + \ldots, \label{g(t)-as}
\end{equation}
where  $\alpha_{k}$, are real numbers for $k=1,2, \ldots$ and $\alpha_{1} > 0$ and the sign of $\alpha_{k}$ for $k >1$ depends on the model considered (see
\cite{Urbanowski:2008kra}).

An important property of $h(t)$ can be found using the relation $| \phi (t) \rangle = \exp\,[-\frac{i}{\hbar}t\mathfrak{H}] | \phi\rangle$, which means that the amplitude ${\cal
A}(t)$ can be written as follows: ${\cal A}(t) \equiv
\langle \phi|\exp\,[-\frac{i}{\hbar}t\mathfrak{H}] | \phi\rangle$.
It is not difficult to see that
this form of ${\cal A}(t)$
and hermiticity of $\mathfrak{H}$ imply that \cite{Fonda}
\begin{equation}
({\cal A}(t))^{\ast} ={\cal A}(-t). \label{amp-ast}
\end{equation}

The conclusion resulting from (\ref{amp-ast})
and from the relation (\ref{h(t)}) is that
\begin{equation}
h(-t) = \left(h(t)\right)^{\ast}. \label{h=h-ast}
\end{equation}
Therefore there must be
\begin{equation}
E(-t) = E(t) \quad \text{and} \quad {\it\Gamma} (-t) = - {\it\Gamma} (t), \label{even+odd}
\end{equation}
That is, the instantaneous energy
$ E(t) = \Re\,[h(t)]$ is an even function of time $t$ and the instantaneous decay rate $ {\it\Gamma}(t) = -2\,\Im\,[h(t)]$ an odd function of $t$.

\section{Calculations and results}

As it was  said in the  previous Section in order to calculate the survival amplitude ${\cal A}(t)$ within
the Fock--Krylov theory of unstable states we need the energy density distribution function $\omega(E)$.
From  an analysis of general properties of the energy (mass)  distribution
functions $\omega (E)$  of real unstable systems it follows that
$\omega (E)$ has properties analogous to the scattering amplitude, i.e.,
it can be
decomposed into a threshold factor, a pole-function $P(E)$ with a simple pole
and a smooth form factor $f(E)$:
There is
\begin{equation}
\omega(E) = {\it\Theta}(E-E_{\rm min})\,(E-E_{\rm min})^{\alpha_{l}}\,P(E)\,f(E), \label{omega}
\end{equation}
where $\alpha_{l}$ depends on the angular momentum $l$ through $\alpha_{l} = \alpha + l$, \cite{Fonda}
(see equation (6.1) in \cite{Fonda}),  $0 \leq \alpha <1$)
 and ${\it\Theta}(E)$ is a step function: ${\it\Theta}(E) = 0\;\;{\rm  for}\;\; E \leq 0$
and ${\it\Theta}(E) = 1\;\;{\rm for}\;\;E>0  $ and $f(E)$ is such a function that $P(E)\,f(E) \to 0$ as $E \to \infty$. The simplest choice is to take $\alpha = 0, l=0, f(E) = 1$ and
to assume that $P(E)$ has a Breit--Wigner (BW) form of the energy distribution density. (The mentioned Breit--Wigner distribution  was found when the cross--section of slow neutrons
was analyzed \cite{Breit:1936zzb}).
It turns out that the decay curves obtained in this simplest case are very similar in form to the curves calculated for
the above described more general $\omega (E)$,
(see \cite{nowakowski,Kelkar2021} and analysis in \cite{Fonda}).
So to find the most typical properties of the decay process it is sufficient to make the relevant calculations for  $\omega (E)$ modeled by the the Breit--Wigner
distribution of the energy density:
\begin{equation}
\begin{split}
\omega (E) &\equiv \omega_{{BW}}(E)  \\
&\stackrel{\text{def}}{=} \frac{N}{2\pi}\, {\it\Theta} (E - E_{{min}}) \
\frac{{\it\Gamma}_{0}}{(E-E_{0})^{2} +
(\frac{{\it\Gamma}_{0}}{2})^{2}}, \label{omega-bw}
\end{split}
\end{equation}
where $N$ is a normalization constant.
The parameters $E_{0}$ and ${\it\Gamma}_{0}$ correspond to the energy of the system in the metastable state and its decay rate at the exponential (or canonical) regime of the decay
process. $E_{{min}}$ is the minimal (the lowest) energy of the system.
For  $\omega (E) = \omega_{BW}(E)$ one can find relatively easy an analytical form of ${\cal A}(t)$ at very late times as well as an  asymptotic analytical form of $h(t)$, $E(t)$ and
${\it\Gamma} (t)$ for such times.
In previous Section  it was stated that $\omega (E)$ contains information characterizing the given metastable state: In the case $\omega (E) = \omega_{BW}(E)$ quantities $E_{0}$,
${\it\Gamma}_{0}$ and $E_{{min}}$ are exactly the parameters characterizing the metastable state considered. The different values of these parameters correspond to different
metastable states.

Inserting $\omega_{{BW}}(E)$ into formula (\ref{a-spec}) for the amplitude ${\cal A}(t)$ and assuming for simplicity that $t_{0}^{init} = 0$, after some algebra one finds that
\begin{equation}
{\cal A}(t) = \frac{N}{2\pi}\,
e^{\textstyle{- \frac{i}{\hbar} E_{0}t }}\,{\cal I}_{\beta}\left(\frac{{{\it\Gamma}}_{0} t}{\hbar}\right) \label{I(t)a}
\end{equation}
where
\begin{equation}
{\cal I}_{\beta}(\tau) \stackrel{\rm def}{=}\int_{-\beta}^{\infty}
 \frac{1}{\eta^{2} + \frac{1}{4}}\, e^{\textstyle{- i\eta\tau}}\,d\eta. \label{I(t)}
\end{equation}
Here $\tau = \frac{{\it\Gamma}_{0}\,t}{\hbar} \equiv \frac{t}{\tau_{0}}$, $\tau_{0}$ is the lifetime, $\tau_{0} = \frac{\hbar}{{\it\Gamma}_{0}}$, and $\beta = \frac{E_{0} -
E_{min}}{{\it\Gamma}_{0}} > 0$.

Having the amplitude ${\cal A}(t)$ we can use it to
analyze properties of the instantaneous energy $E(t)$ and instantaneous decay rate ${\it\Gamma} (t)$.
These quantities are defined using the effective Hamiltonian $h(t)$ which is build from ${\cal A}(t)$. In order to find $h(t)$ we need  the quantity $i\,\hbar\, \frac{\partial {\cal
A}(t)}{\partial t}$ (see (\ref{h(t)})). From Eq.~(\ref{I(t)a}) one finds that
\begin{equation}
i \hbar \frac{\partial {\cal A}(t)}{\partial t} = E_{0} \,{\cal A}(t)
+ {\it\Gamma}_{0}\,\frac{N}{2\pi}\,e^{\textstyle{-\frac{i}{\hbar} E_{0}t}}\,{\cal J}_{\beta}(\tau(t)), \label{J-R-1}
\end{equation}
where
\begin{equation}
{\cal J}_{\beta}(\tau) = \int_{- \beta}^{\infty}\,\frac{x}{x^{2} + \frac{1}{4}}\,e^{\textstyle{-ix\tau}}\,dx, \label{J-R}
\end{equation}
or simply (see (\ref{I(t)})),
\begin{equation}
{\cal J}_{\beta}(\tau) \equiv i\frac{\partial {\cal I}_{\beta} (\tau)}{\partial \tau}. \label{J-R-eq}
\end{equation}

Now the use of (\ref{I(t)a}), (\ref{J-R-1}) and (\ref{h(t)}) leads to the conclusion that within the model considered there is,
\begin{equation}
h(t) = i \hbar \frac{1}{{\cal A}(t)}\,\frac{\partial {\cal A}(t)}{\partial t} = E_{0} + {\it\Gamma}_{0}\,\frac{{\cal J}_{\beta}(\tau(t))}{{\cal I}_{\beta}(\tau(t))}, \label{h(t)-1}
\end{equation}
which means that
\begin{equation}
E(t) = \Re\,[h(t)] = E_{0} + {\it\Gamma}_{0}\,\Re\,\left[\frac{{\cal J}_{\beta}(\tau(t))}{{\cal I}_{\beta}(\tau(t))}\right], \label{E(t)-2}
\end{equation}
and
\begin{equation}
{\it\Gamma} (t) = -2\,\Im[h(t)] = -2\,{\it\Gamma}_{0}\,\Im\left[\frac{{\cal J}_{\beta}(\tau(t))}{{\cal I}_{\beta}(\tau(t))}\right]. \label{g(t)-a}
\end{equation}

Using (\ref{I(t)a}) --- (\ref{g(t)-a}) one can find  analytically within the model considered late time asymptotic forms of $E(t)$ and ${\it\Gamma}(t)$. There is for $t \to \infty$
(see \cite{jcap-2020,R-U}):
\begin{equation}
 \label{Re-h-as}
 \begin{split}
{E(t)\, \vline}_{\,t \rightarrow \infty} &= {\Re\,[h(t)] \, \vline}_{t \to \infty}\\
&\simeq { E}_{{min}}\, -\,2\,
\frac{ \beta }{{\it\Gamma}_{0}\,(\beta^{2} + \frac{1}{4}) } \,
\left(\frac{\hbar}{t} \right)^{2} + \ldots ,
\end{split}
\end{equation}
and,
\begin{equation} \label{Im-h-as}
{{\it\Gamma}(t)\, \vline}_{\,t \rightarrow \infty} = - 2 \Im\,[h(t)]
\simeq 2\,\frac{\hbar}{t}  + \ldots \, .
\end{equation}

\begin{figure}[H]
\begin{center}
\subfigure[\scriptsize{ The survival probability ${\cal P}(\tau)$}]{
\includegraphics[width=68mm]{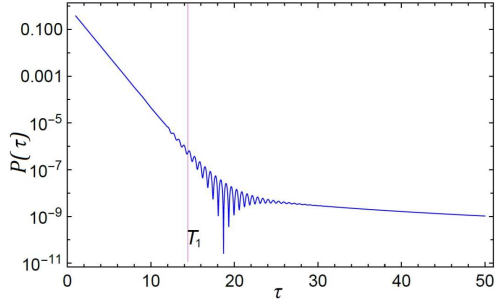}
\label{fg1}
}
\;\;
\subfigure[\scriptsize{$\kappa(\tau)$}]{
\includegraphics[width=65mm]{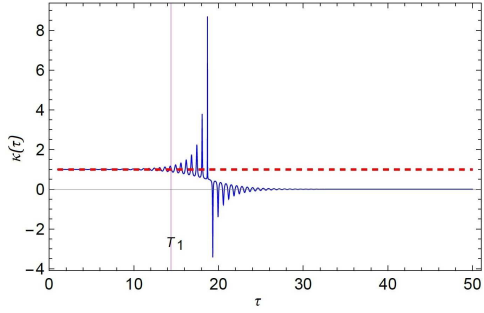}
\label{fg2}
}
\\
\subfigure[\scriptsize{The ratio $E(\tau)/E_{min}$}]{
\includegraphics[width=68mm]{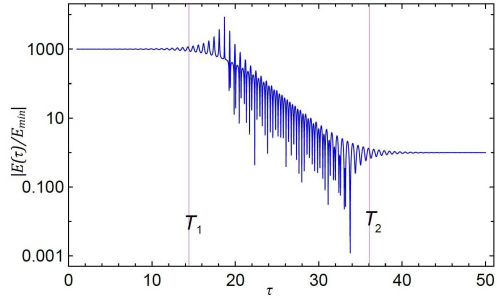}
\label{fg3}
}
\end{center}
\caption{\small{Results obtained for  $\omega_{BW}(E)$ given by Eq. (\ref{omega-bw}). The case $\beta =10$. Fig (\ref{fg1}) ---  A  decay curve ${\cal P}(\tau) = |{\cal
A}(\tau)|^{2}$; Fig (\ref{fg2}) ---  An illustration of the typical behavior of energy $E(\tau)$: The solid line --- $\kappa (\tau )\equiv \left(E(\tau) - E_{min}\right)/\left(E_{0} -
E_{min}\right)$, The dashed line  ---  $E(\tau) = E_{0} = \text{const}$ ($\kappa (\tau) = 1$);
Fig (\ref{fg3}) --- The modulus of the ratio $E(\tau)/E_{min}$, the case  $E_{0}/E_{min} = 1000$.  In all figures the time, $t$,  is measured in lifetimes $\tau_{0}$:  $\tau  =t /
\tau_{0} $ and    $\tau_{0} = \hbar/{\it\Gamma}_{0}$ is the lifetime.}}
\label{fg-all}
\end{figure}

In order to visualize properties of $E(t)$ it is convenient to use the following function
\begin{equation}
\kappa (t) \stackrel{\rm def}{=} \frac{E(t) - E_{\text{min}}}{E_{0} - E_{\text{min}}} . \label{kappa}
\end{equation}
Using (\ref{E(t)-2}) one finds that
\begin{equation}
E(t) - E_{{min}} = E_{0} - E_{{min}} + \Gamma_{0}\,\Re\,\Big[\frac{{\cal J}_{\beta}(\tau)}{{\cal I}_{\beta}(\tau)}\Big], \label{E(t)-1}
\end{equation}
Now, it to divide two sides of equation (\ref{E(t)-1}) by  \linebreak $(E_{0} - E_{{min}})$ then one obtains the function $\kappa (t)$ (see (\ref{kappa})) we are looking for
\begin{equation}
\kappa (\tau(t)) = 1 + \frac{1}{\beta}\,\Re\,\Big[\frac{{\cal J}_{\beta}(\tau(t))}{{\cal I}_{\beta}(\tau(t))}\Big]. \label{kappa-1}
\end{equation}
Using the above derived formulae  one can find  numerically within the model considered the survival probability ${\cal P}(t)$ and $\kappa(\tau(t))$ describing the behavior of the
instantaneous energy $E(t)$. Results of these calculations performed for chosen $\beta$ are presented in Figs (\ref{fg1}), (\ref{fg2}).

\begin{figure}[H]
\subfigure[\scriptsize{ $y$--axis -- Normal scale }]{
\includegraphics[width=65mm]{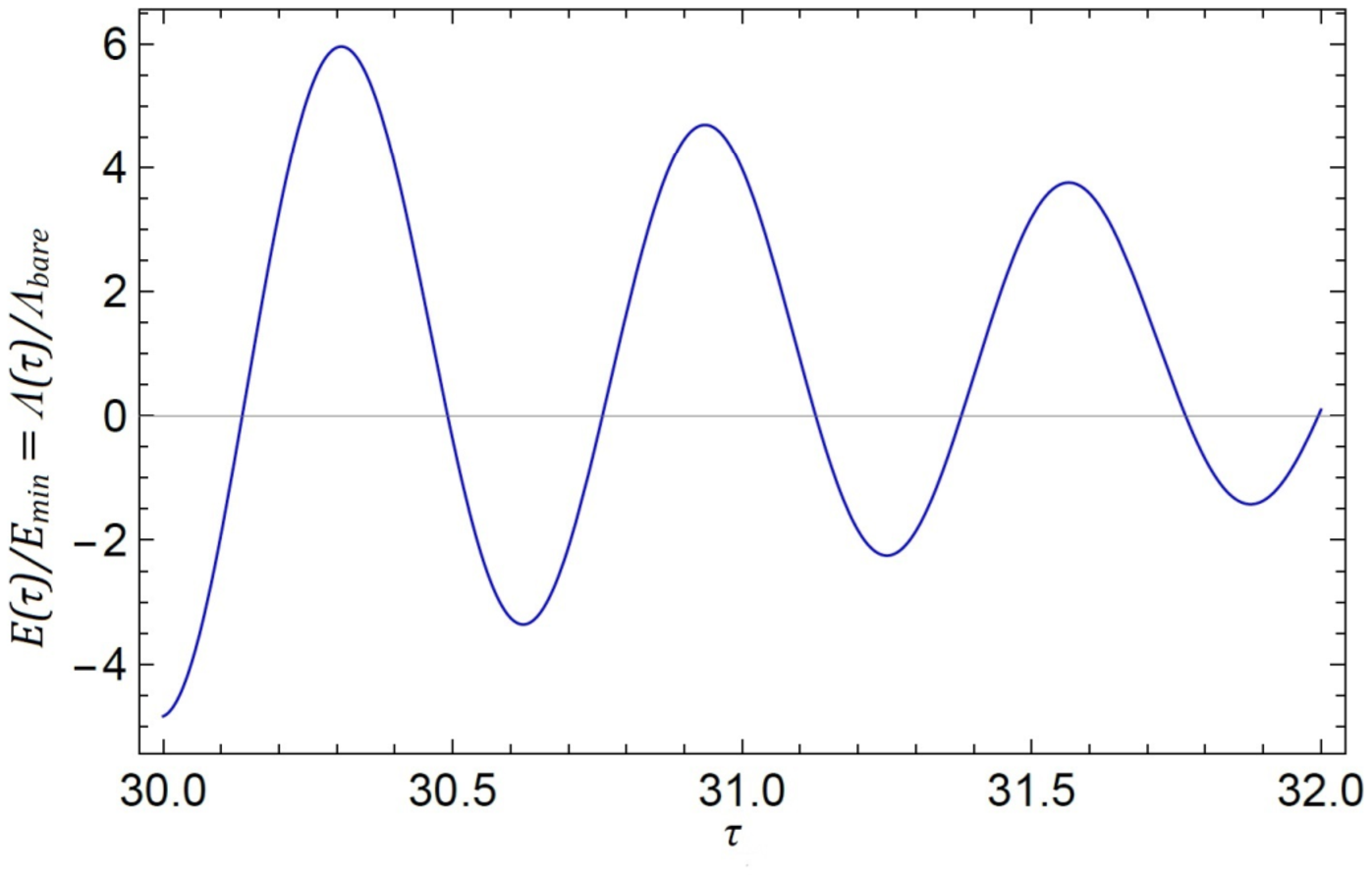}
\label{f1a}
}
\;
\subfigure[\scriptsize{$y$--axis -- Logarithmic scale: Modulus of the ratio $E(\tau)/E_{min}$ }]{
\includegraphics[width=68mm]{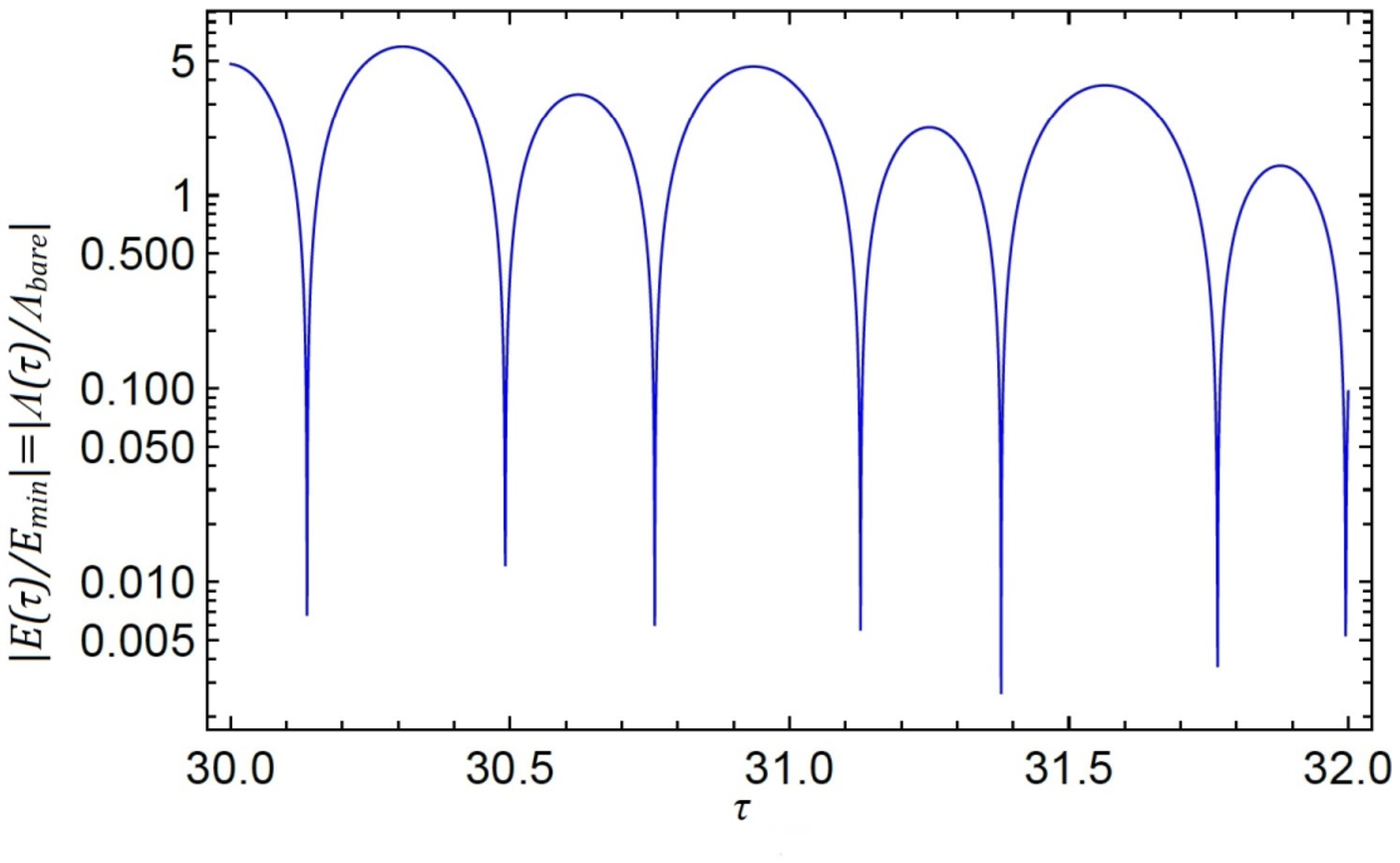}
\label{f1b}
}
\caption{\small The  ratio $E(\tau)/E_{min}$ (An enlarged  part of Fig (\ref{fg3})). The case $\beta = 10$ and $E_{0}/E_{min} = \Lambda_{0}/\Lambda_{bare} = 1000$.
In all figures the time $t$ is measured in lifetimes: $\tau = t/\tau_{0}$: $\tau = t/\tau_{0}$ and $\tau_{0}  = \hbar/{\it\Gamma}_{0}$.
}
\label{f1-all}
\end{figure}

\begin{figure}[H]
\centering
\subfigure[\scriptsize{The case $\beta = 10$.}]{
\includegraphics[width=60mm]{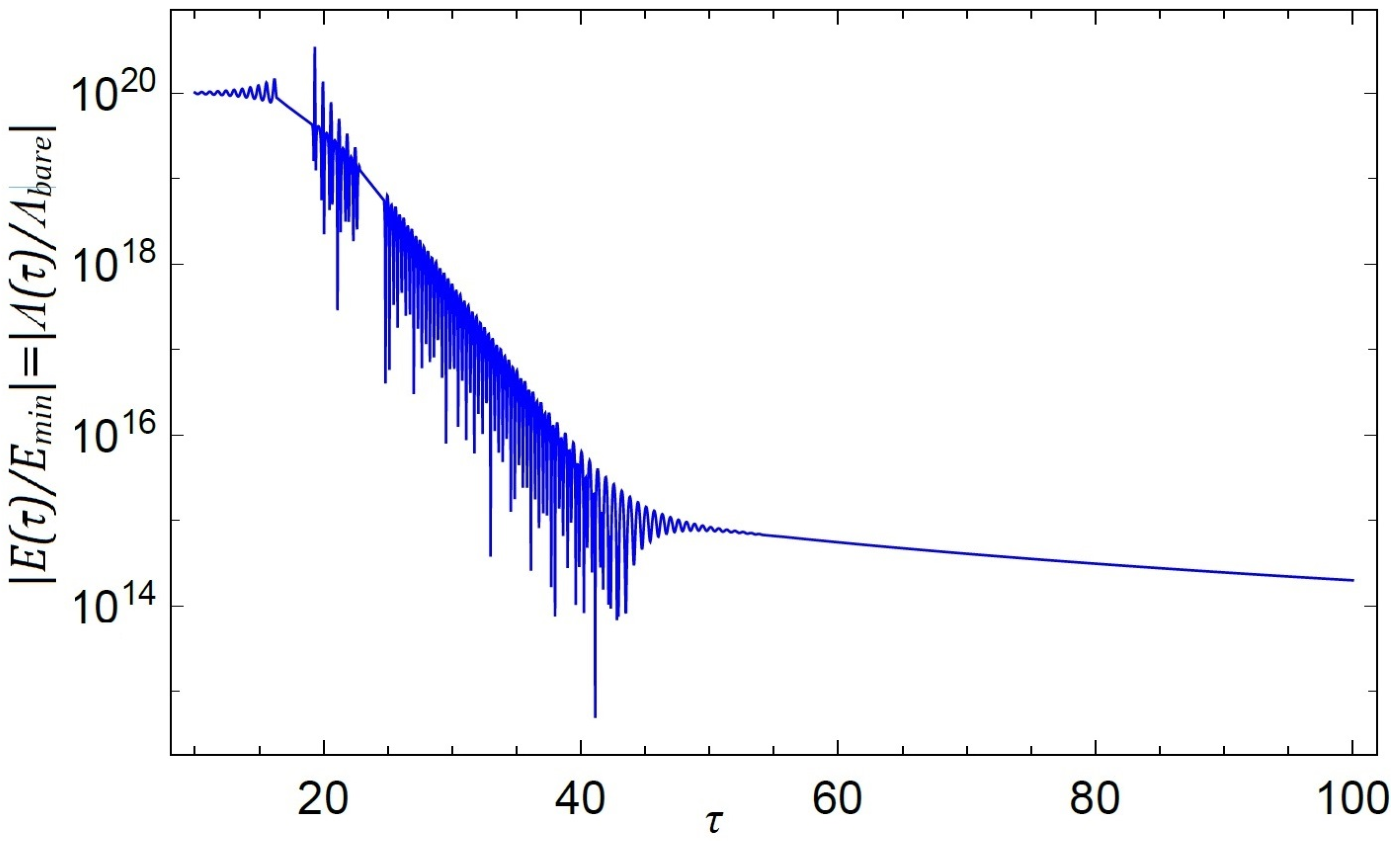}
\label{f21}
}
\;
\subfigure[\scriptsize{The case $\beta= 100$.}]{
\includegraphics[width=60mm]{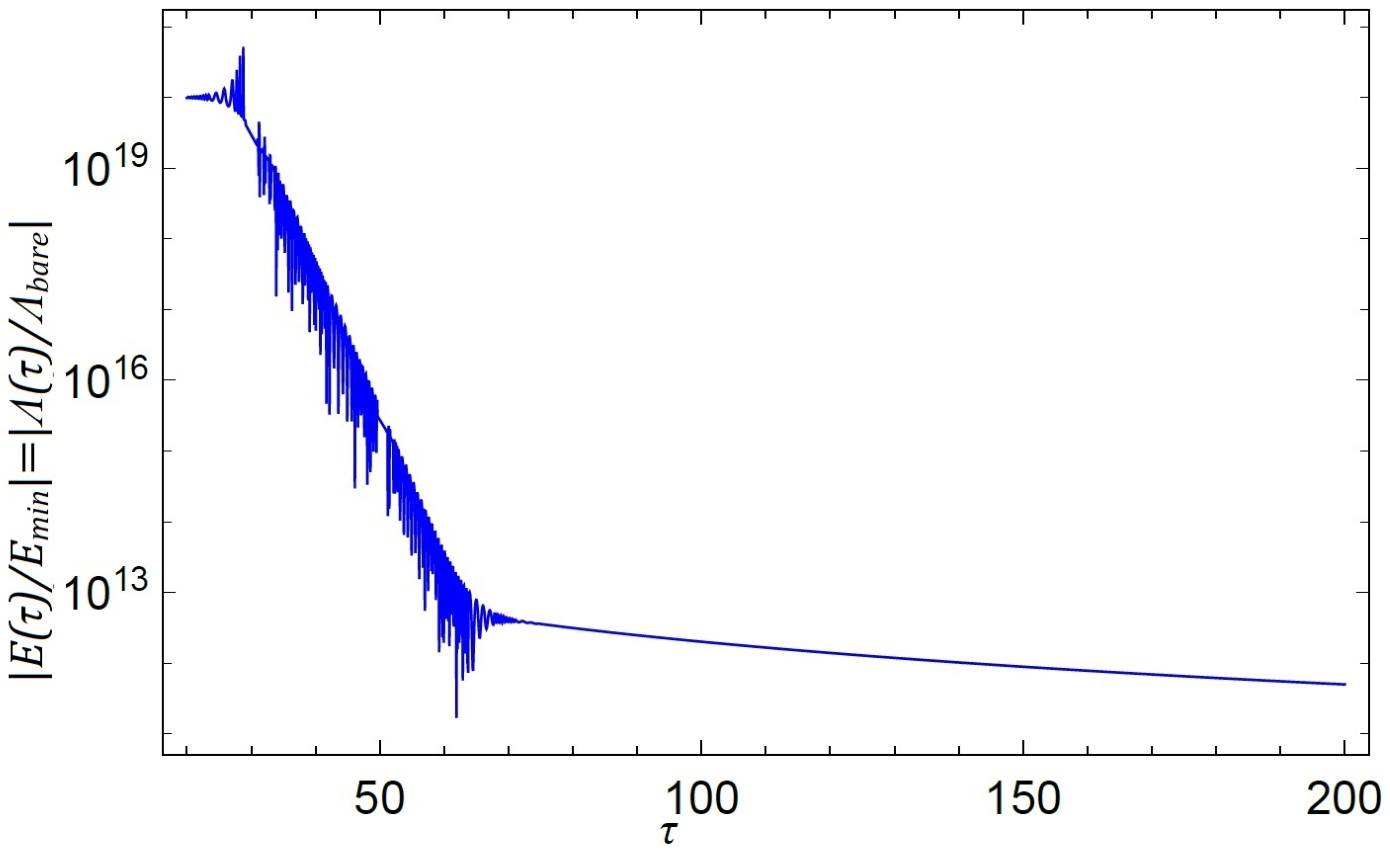}
\label{f22}
} \\

\subfigure[\scriptsize{The case $\beta = 1000$.}]{
\includegraphics[width=60mm]{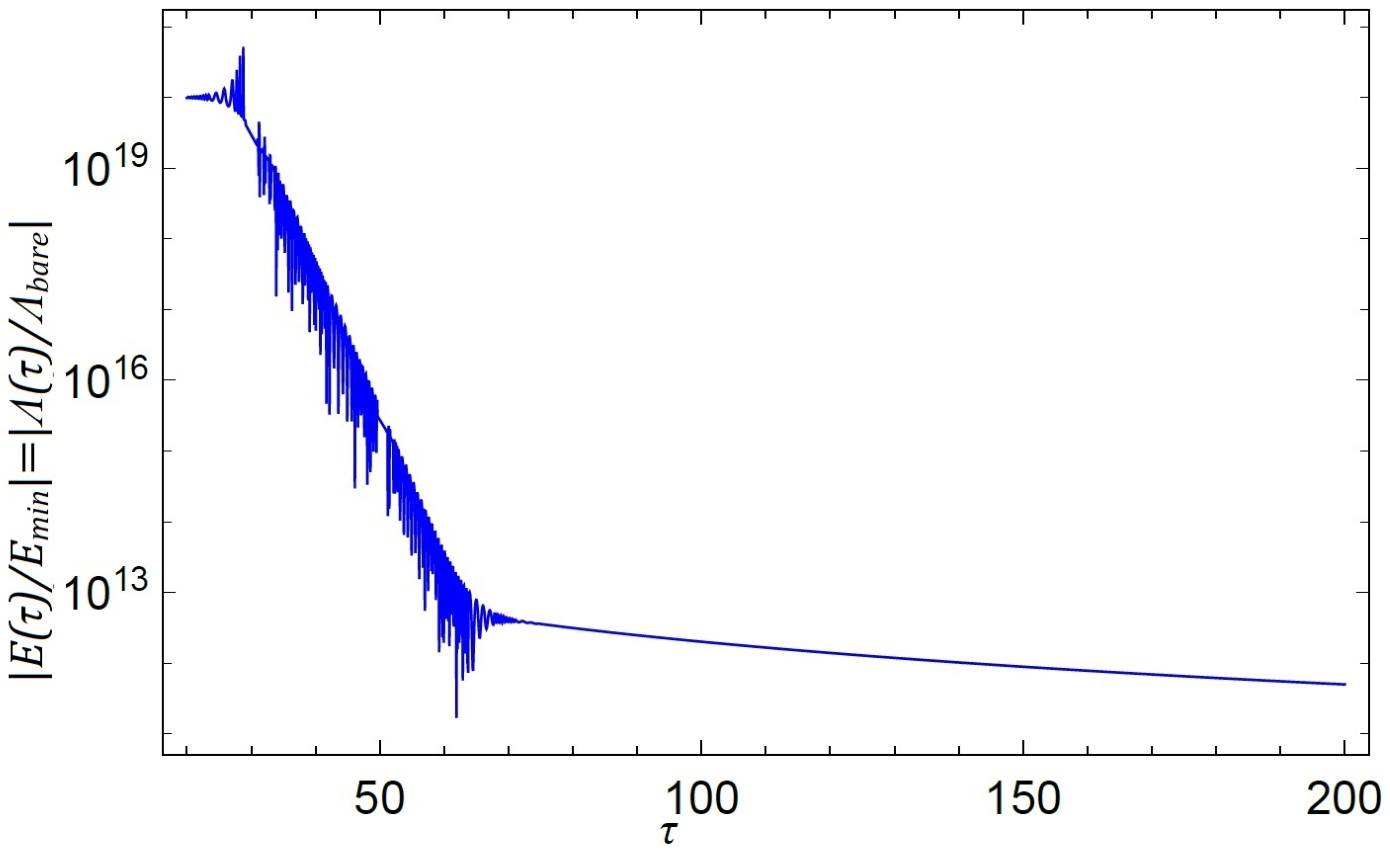}
\label{f23}
}
\;
\subfigure[\scriptsize{The case $\beta= 10000$.}]{
\includegraphics[width=60mm]{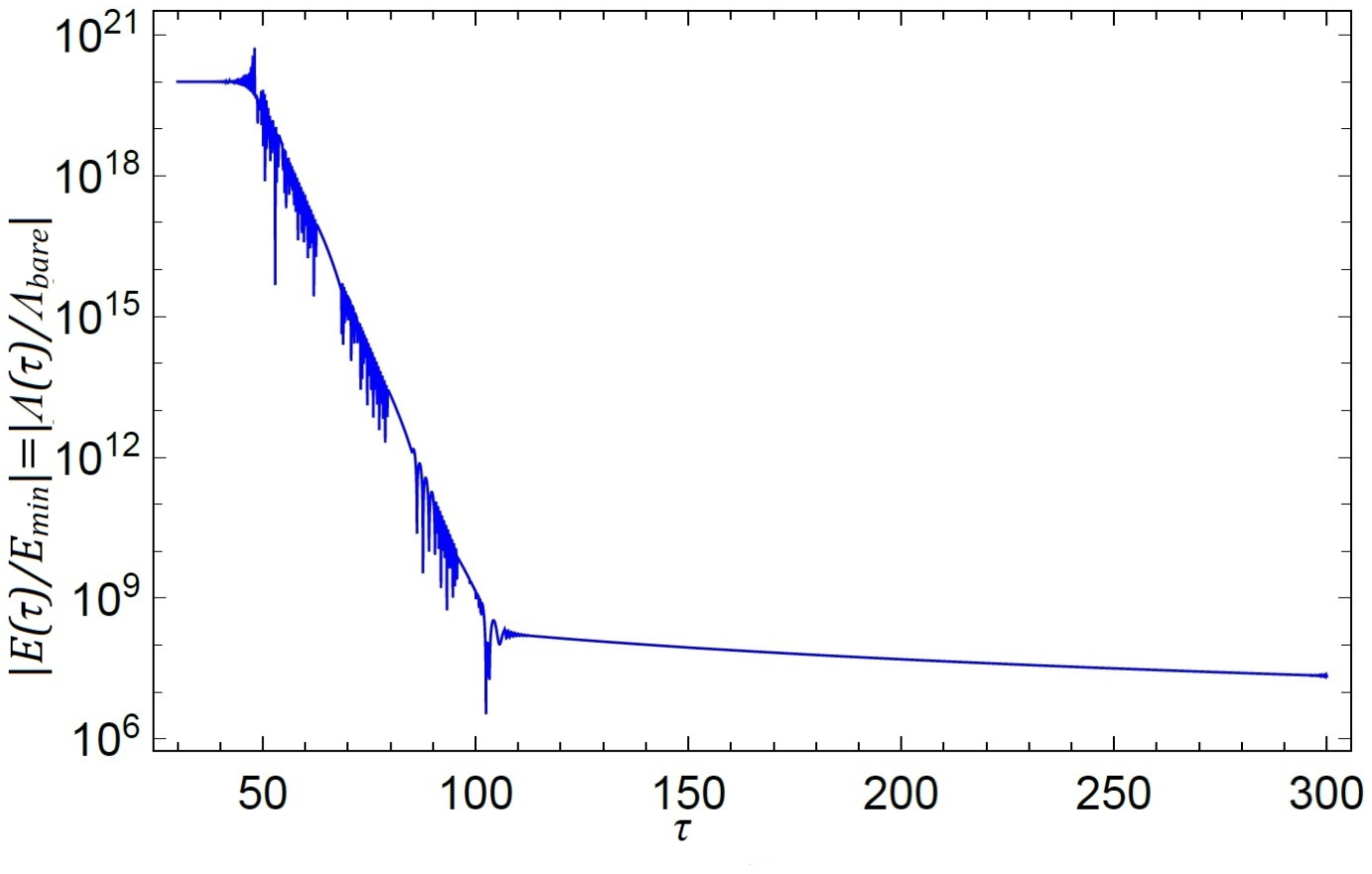}
\label{f24}
}
\caption{\small   The ratio $E(\tau)/E_{min}  \equiv \Lambda (\tau)/\Lambda_{bare}$ obtained for a $\omega_{BW}(E)$ given by formula (\ref{omega-bw}).
In all figures the time $t$ is measured in lifetimes: $\tau = t/\tau_{0}$: $\tau = t/\tau_{0}$ and $\tau_{0}  = \hbar/{\it\Gamma}_{0}$.
The case $E_{0}/E_{min} =  \Lambda_{0}/\Lambda_{bare}= 10^{20}$.
}
\label{f2-all}
\end{figure}

\begin{figure}[h!]
\centering
\subfigure[\scriptsize{The case $E_{0}/E_{min} = \Lambda_{0}/\Lambda_{min} = 10^{30}$.}]{
\includegraphics[width=60mm]{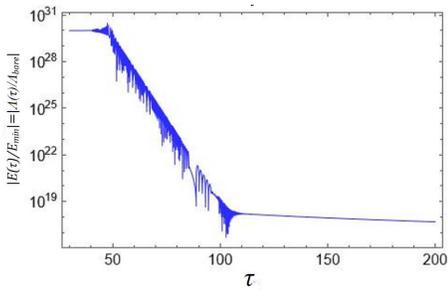}
\label{f2a1}
}
\;
\subfigure[\scriptsize{The case $E_{0}/E_{min} = \Lambda_{0}/\Lambda_{min} = 10^{40}$}]{
\includegraphics[width=60mm]{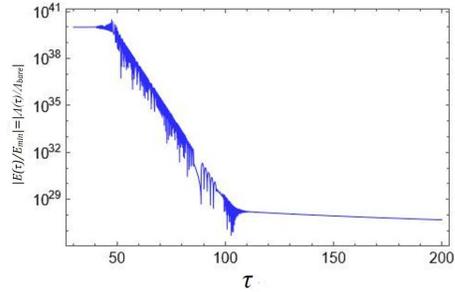}
\label{f2a2}
}
\caption{\small   The ratio $E(\tau)/E_{min}  \equiv \Lambda (\tau)/\Lambda_{bare}$ obtained for a $\omega_{BW}(E)$ given by formula (\ref{omega-bw}).
In all figures the time $t$ is measured in lifetimes: $\tau = t/\tau_{0}$: $\tau = t/\tau_{0}$ and $\tau_{0}  = \hbar/{\it\Gamma}_{0}$.
The case $\beta= 10000$.
}
\label{f2a-all}
\end{figure}

From the results of the previous Section and those presented above and also in Fig (\ref{fg2})  it follows that
a behavior of the instantaneous energy $E(t)$ and the instantaneous decay rate ${\it\Gamma}(t)$ differ
depending on the time domain in which we examine their values. At canonical decay times $0 \ll t < T_{1}$ they are  close to a good approximation to the values resulting from the
Weisskopf--Wigner theory of spontaneous emission. For times $t >T_{2} $ their behavior is described by late time asymptotic formulae (\ref{E(t)-as}) and (\ref{g(t)-as}). At transition
time region of times $T_{1} < t < T_{2}$ the instantaneous energy $E(t)$ and the instantaneous decay rate ${\it\Gamma}(t)$ decrease to their late time asymptotic forms.
Unfortunately the information of how fast  $E(t)$ tends to its asymptotic form (\ref{E(t)-as}) at times $t > T_{1}$
is invisible in $\kappa(\tau(t))$ --- see Fig (\ref{fg2}).  In order to remove this   deficiency one should come back to the Eq. (\ref{kappa}).
Namely by manipulating Eqs (\ref{kappa}) --- (\ref{kappa-1})  we get the desired result,
\begin{equation}
\frac{E(\tau(t))}{E_{{min}}} = 1 + \left(\frac{E_{0}}{E_{{min}}} - 1\right)\,\kappa(\tau(t)). \label{E(t)-Emin}
\end{equation}
This last equation can be used to show how fast $E(t)$ tends to its asymptotic form (\ref{E(t)-as}) at times $t > T_{1}$ for assumed $\beta$ and the ratio $\frac{E_{0}}{E_{{min}}}$.
Results obtained for $\omega(E) = \omega_{BW}(E)$ are presented in graphical form in Figs (\ref{fg-all}), (\ref{f1-all}), (\ref{f2-all}) and Fig (\ref{f2a-all}).

We show now that similar results can be obtained
not only in
the approximate case $\omega_{BW}(E)$
of the density  of the energy (mass) distribution $\omega(E)$
but also when one considers  a more general forms of $\omega (E)$.
As it was mentioned  $\omega (E)$
should have
a form given by 
Eq. (\ref{omega}), where the simple pole contribution, $P(E)$, is often modeled by $\omega_{BW}(E)$.

\begin{figure}[H]
\centering
\subfigure[\scriptsize{The case $\beta = 10$, $\eta=0.25$.}]{
\includegraphics[width=60mm]{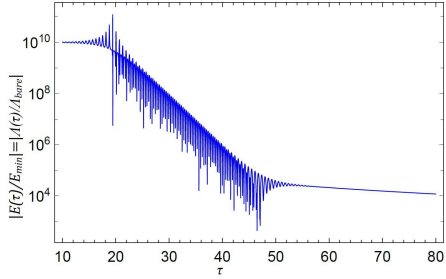}
\label{f31}
}
\qquad
\subfigure[\scriptsize{The case $\beta= 10$, $\eta=0.5$}]{
\includegraphics[width=60mm]{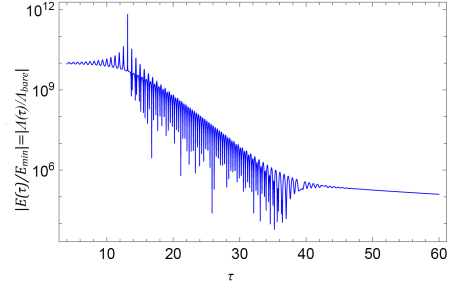}
\label{f32}
}
\caption{\small  A ratio $E(\tau)/E_{min} \equiv \Lambda (\tau)/\Lambda_{bare}$ obtained for a $\omega_{BW}(E)$ given by formula (\ref{omega-bw}).
In all figures the time $t$ is measured in lifetimes: $\tau = t/\tau_{0}$: $\tau = t/\tau_{0}$ and $\tau_{0}  = \hbar/{\it\Gamma}_{0}$.
The case $E_{0}/E_{min} = \Lambda_{0}/\Lambda_{bare} = 10^{10}$.}
\label{f3-all}
\end{figure}

\begin{figure}[h!]
\centering
\subfigure[\scriptsize{The case $\beta = 10000$, $\eta=0.0001$, ${E_{0}}/{E_{min}} ={\Lambda_{0}}/{\Lambda_{bare}}=  10^{14}$.}]{
\includegraphics[width=60mm]{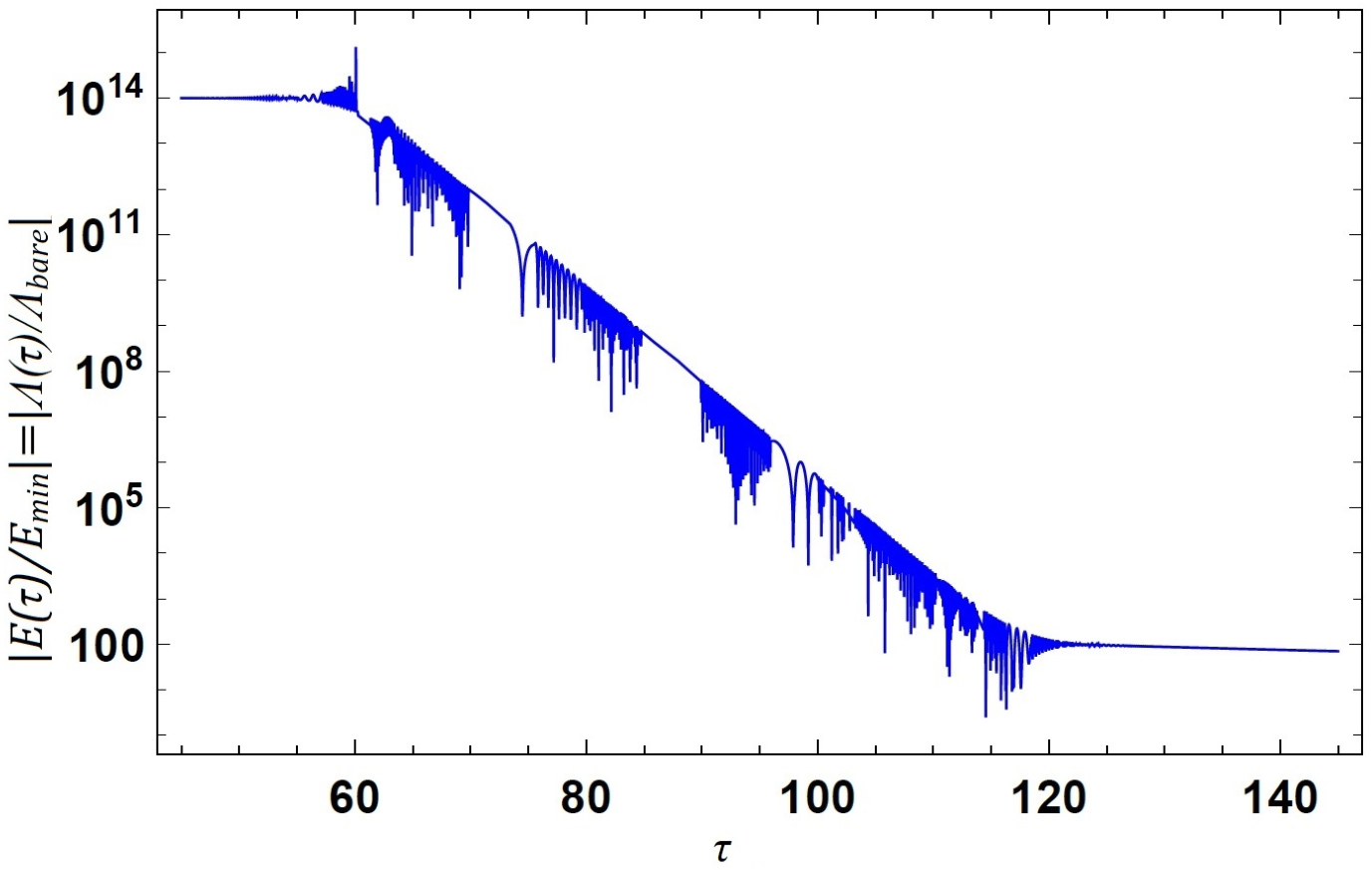}
\label{f71}
}
\;
\subfigure[\scriptsize{The case $\beta= 100$, $\eta=0.1$, ${E_{0}}/{E_{min}} ={\Lambda_{0}}/{\Lambda_{bare}}=  10^{20}$.}]{
\includegraphics[width=60mm]{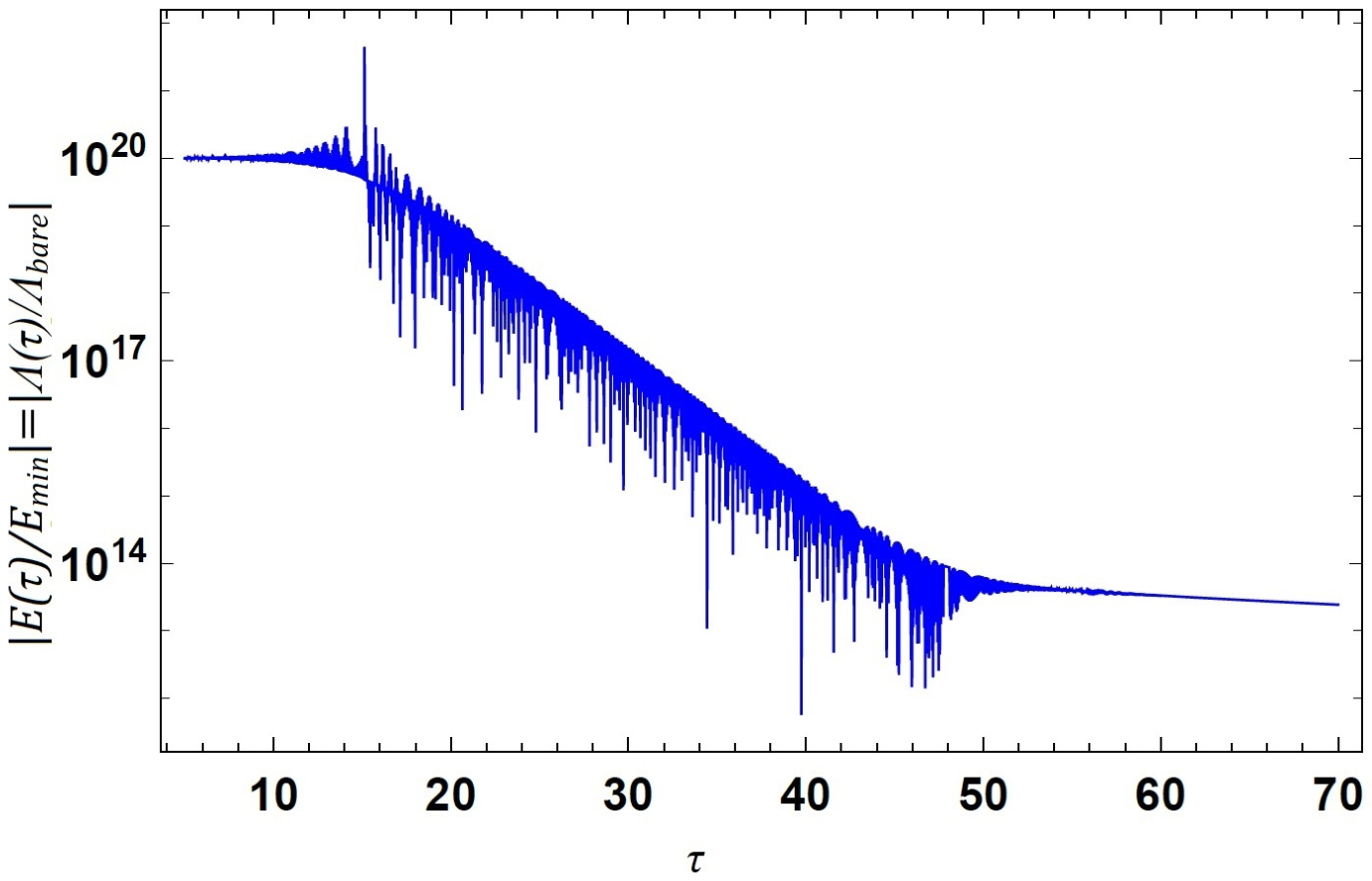}
\label{f72}
}
\caption{\small   A ratio $E(\tau)/E_{min} \equiv \Lambda (t)/\Lambda_{bare}$ obtained for a $\omega(E)$ given by formula (\ref{omega-exp}).
In all figures the time $t$ is measured in lifetimes: $\tau = t/\tau_{0}$: $\tau = t/\tau_{0}$ and $\tau_{0}  = \hbar/{\it\Gamma}_{0}$.}
\label{f7-all}
\end{figure}

Guided by this observation we follow \cite{nowakowski,Kelkar2021} and assume that
\begin{equation}
\begin{split}
\omega_{\eta}(E) &= N\,\sqrt{E - E_{min}}\,\times \\ &\times\frac{{\sqrt{ {\it\Gamma}_{0} }}}{({ E}-{ E}_{0})^{2} +
({{\it\Gamma}_{0}} / {2})^{2}}\,e^{\textstyle{-\eta\,\frac{E - E_{0}+E_{min}}{\it\Gamma_{0}}}},
\label{omega-exp}
\end{split}
\end{equation}
with $\eta >0$.
Inserting this $\omega_{\eta} (E)$ into (\ref{a-spec}) we can calculate the survival amplitude ${\cal A}(t)$ for this case and then the effective Hamiltonian $h(t)$ and thus $E(t)$
that we want to study. Analogously to the above analyzed case of the Breit--Wigner energy density distribution,  $\omega_{BW}(E)$, we get
\begin{equation}
\frac{E(\tau) - E_{min} }{E_{0} - E_{min}}  = \kappa_{\eta}(\tau), \label{kappa-eta}
\end{equation}
where
\begin{equation}
\kappa_{\eta}(\tau) =
 1 +
 \; + \frac{1}{\beta} \, \Re\left[ \frac{{\cal J}_{\beta}^{\eta}(\tau) }{  {\cal I}_{\beta}^{\eta}(\tau) }\; \right]. \label{kappa-eta1}
\end{equation}
and
\begin{eqnarray}
{\cal J}_{\beta}^{\eta}&=& \int_{-\beta}^{\infty} \frac{x \sqrt{x +\beta} }{ {x}^{2} + \frac{1}{4} }\,e^{\textstyle{-x\,\eta}}\,e^{\textstyle{-i x \tau}} \,d x, \label{J-eta}\\
{\cal I}_{\beta}^{\eta} &=& \int_{-\beta}^{\infty} \frac{ \sqrt{x +\beta} }{ {x}^{2} + \frac{1}{4} } \,e^{\textstyle{-x\,\eta}}\,e^{\textstyle{-i x \tau}}\,dx. \label{I-eta}
\end{eqnarray}
After some algebra, analogously to the result (\ref{E(t)-Emin}), we obtain the following relation in the considered here case:
\begin{equation}
\frac{E(\tau(t))}{E_{{min}}} = 1 + \left(\frac{E_{0}}{E_{{min}}} - 1\right)\,\kappa_{\eta}(\tau(t)). \label{E(t)-Emin-eta}
\end{equation}
This equation, similarly to the Eq (\ref{E(t)-Emin}),
can be used to show how fast $E(t)$ tends to its asymptotic form (\ref{E(t)-as}) at times $t > T_{1}$ for assumed $\omega(E)$, and for chosen  $\beta$, $\eta$ and the ratio
$\frac{E_{0}}{E_{{min}}}$. Results obtained for $\omega(E) = \omega_{\eta}(E)$  are presented graphically in Fig (\ref{f3-all}) --- Fig (\ref{f7-all}).

 \begin{figure}[h!]
\centering
\subfigure[\scriptsize{The case $\beta = 100$, $\eta=0.01$.}]{
\includegraphics[width=60mm]{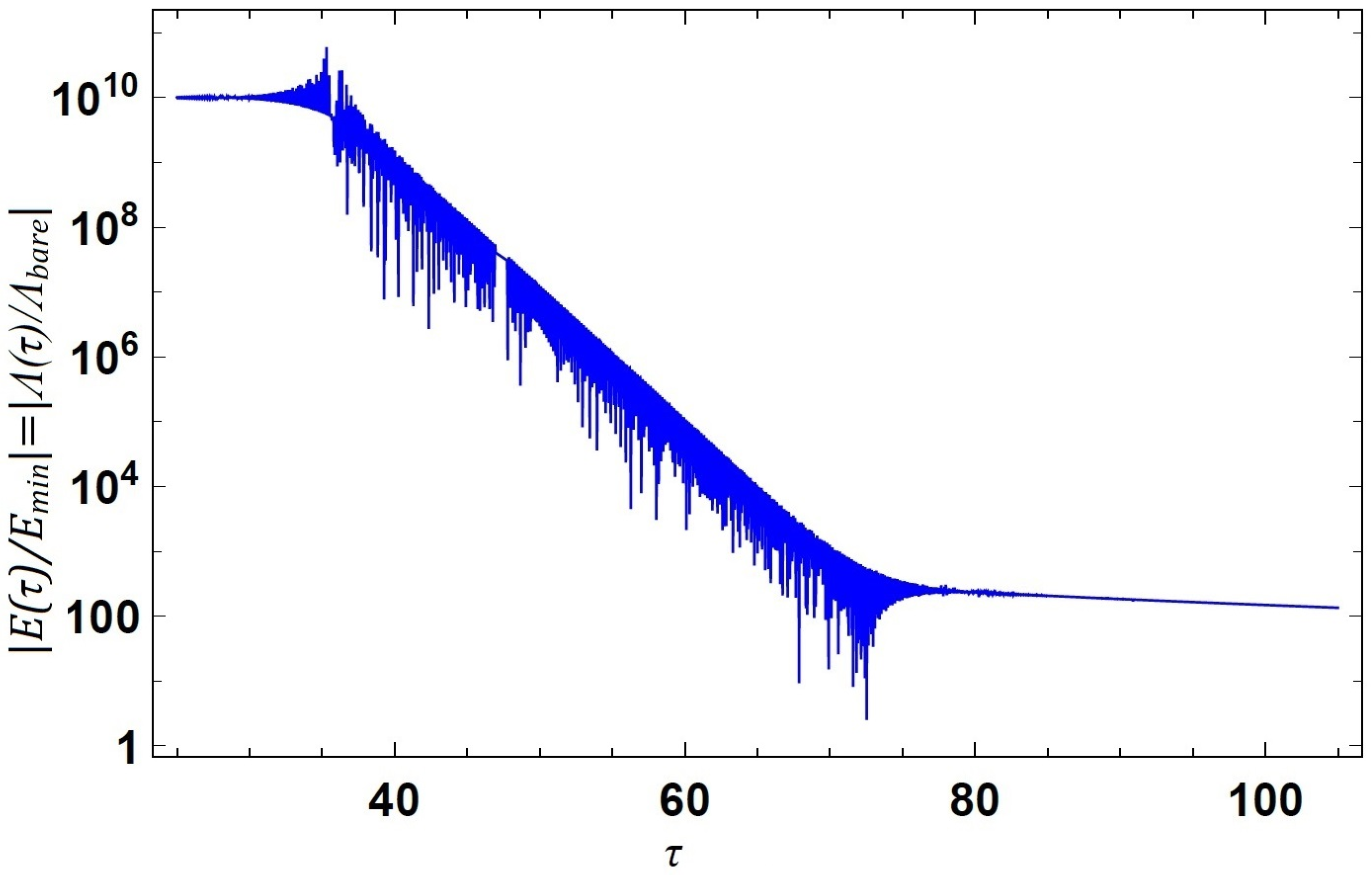}
\label{f41}
}
\qquad
\subfigure[\scriptsize{The case $\beta= 100$, $\eta=0.05$. }]{
\includegraphics[width=60mm]{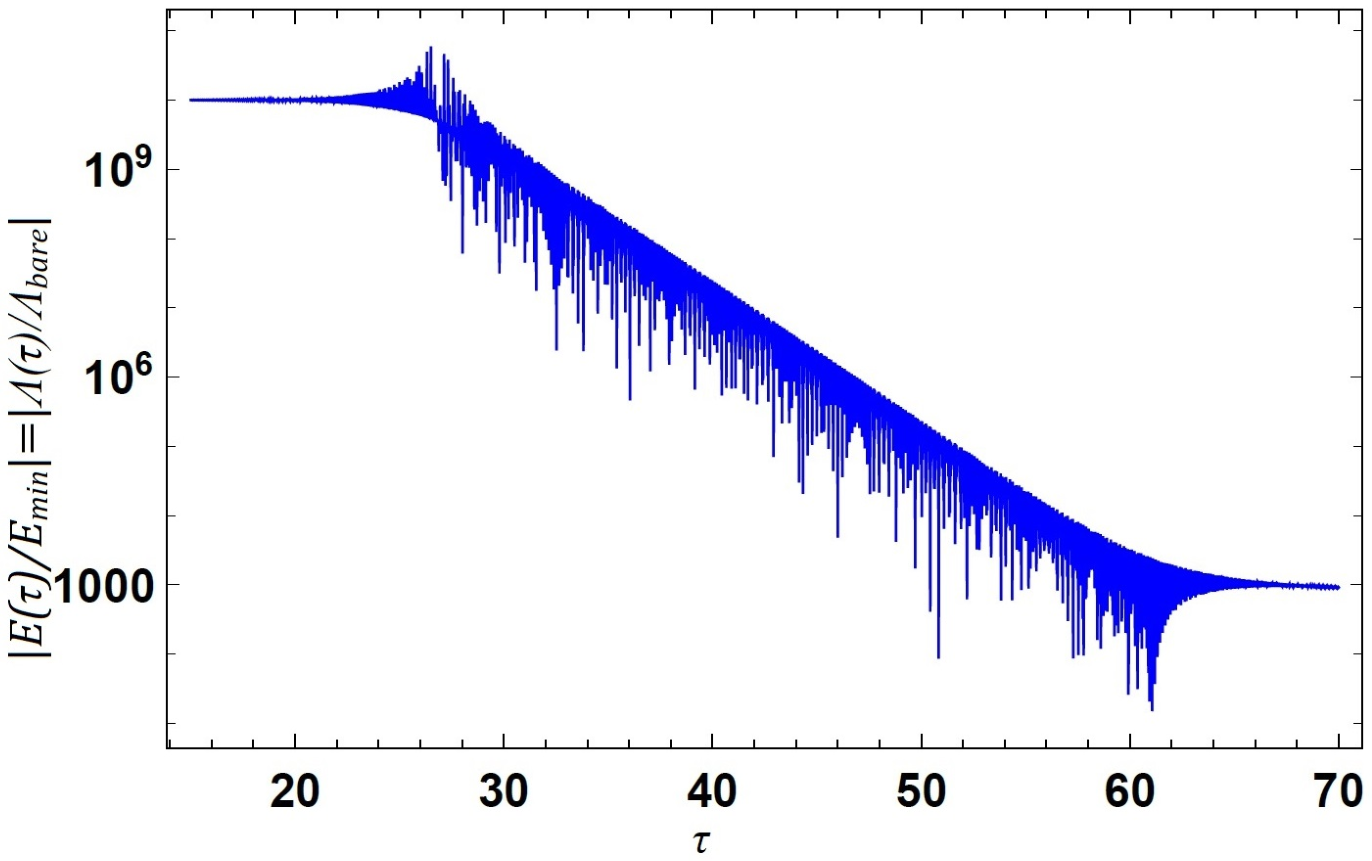}
\label{f42}
} \\
\subfigure[\scriptsize{The case $\beta = 100$, $\eta=0.1$}]{
\includegraphics[width=60mm]{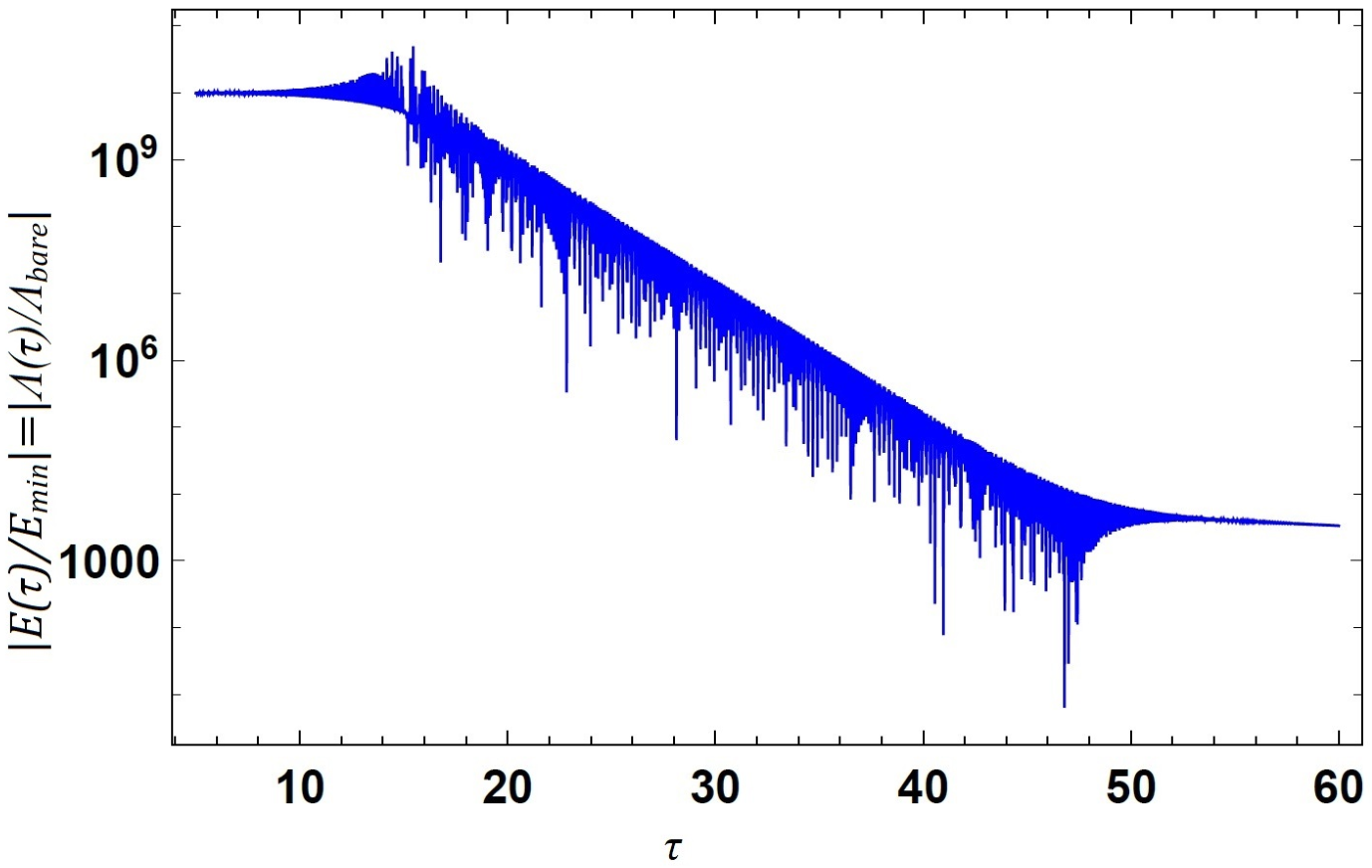}
\label{f43}
}
\caption{\small  A ratio $E(\tau)/E_{min}  \equiv \Lambda (\tau) / \Lambda_{bare}$ obtained for a $\omega_{BW}(E)$ given by formula (\ref{omega-bw}).
The case $E_{0}/E_{min} = \Lambda_{0}/\Lambda_{bare} = 10^{10}$.
In all figures the time $t$ is measured in lifetimes: $\tau = t/\tau_{0}$: $\tau = t/\tau_{0}$ and $\tau_{0}  = \hbar/{\it\Gamma}_{0}$.
}
\label{f4-all}
\end{figure}

\begin{figure}[h!]
\centering
\subfigure[\scriptsize{The case $\beta = 1000$, $\eta=0.01$.}]{
\includegraphics[width=60mm]{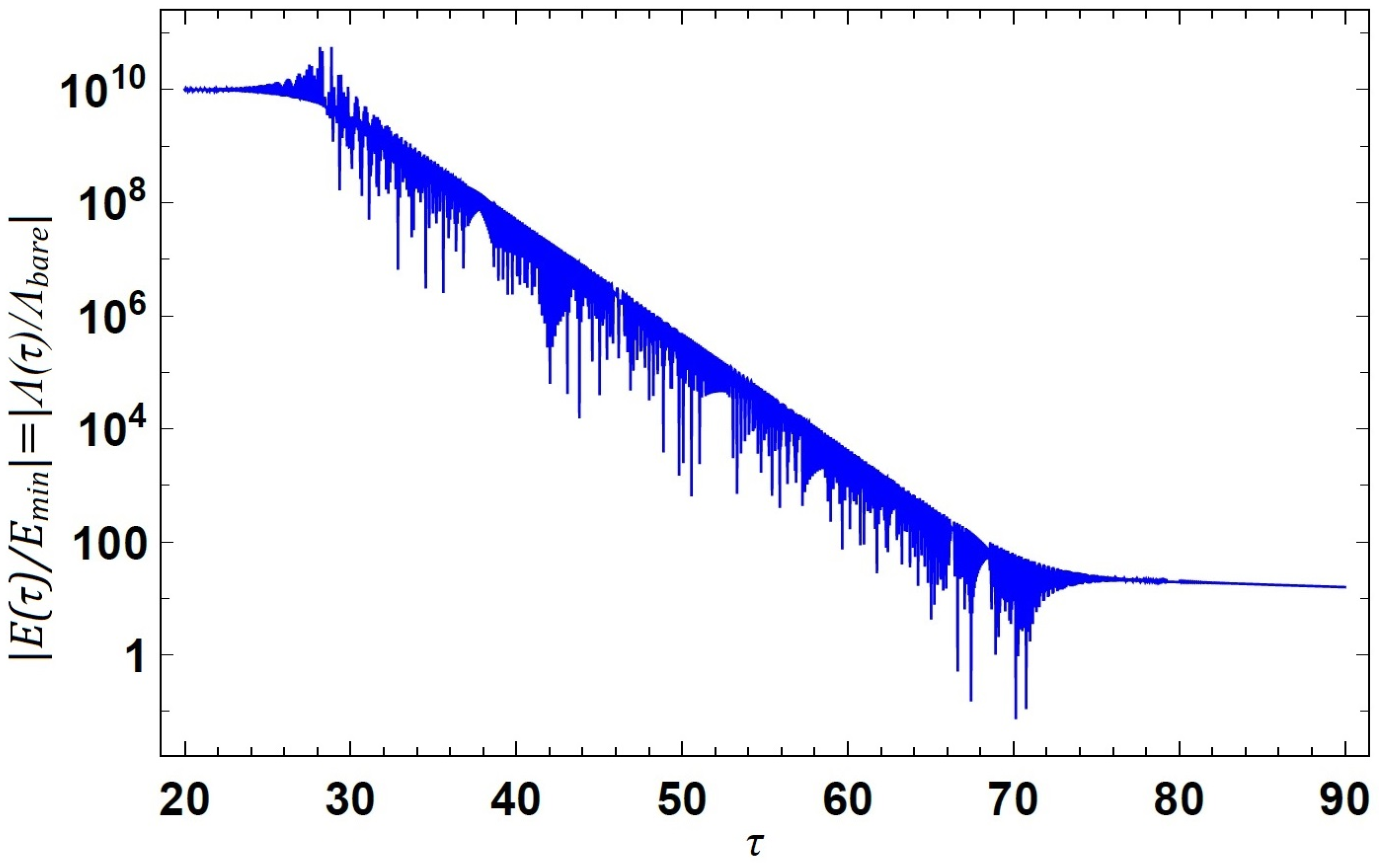}
\label{f51}
}
\;
\subfigure[\scriptsize{The case $\beta= 10000$, $\eta=0.001$.}]{
\includegraphics[width=60mm]{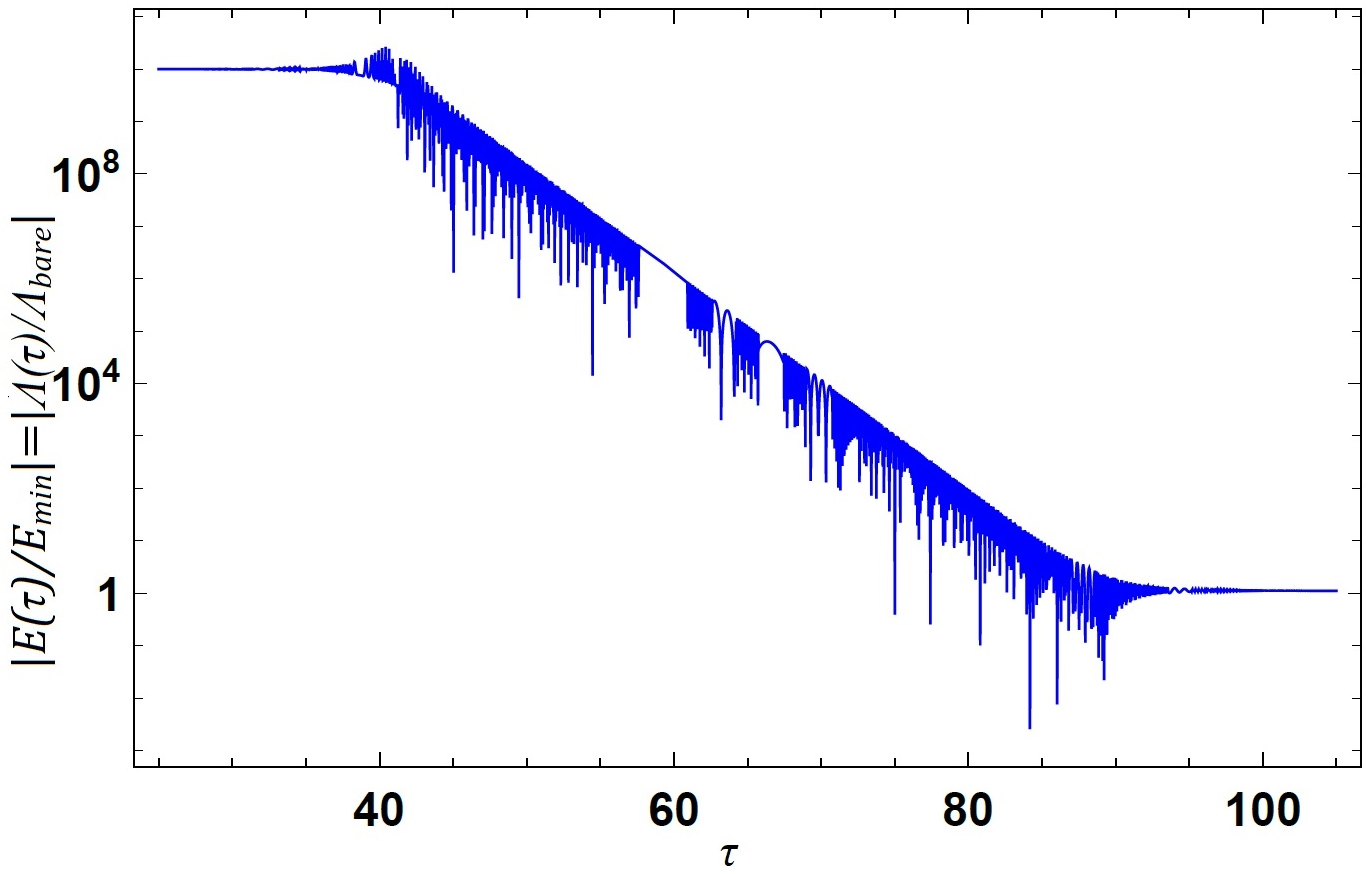}
\label{f52}
}
\caption{ \small  A ratio $E(\tau)/E_{min} \equiv \Lambda (\tau)/\Lambda_{bare}$ obtained for a $\omega_(E)$ given by formula (\ref{omega-exp}).
The case $E_{0}/E_{min}=\Lambda_{0}/\Lambda_{bare}= 10^{10}$.
In all figures the time $t$ is measured in lifetimes: $\tau = t/\tau_{0}$: $\tau = t/\tau_{0}$ and $\tau_{0}  = \hbar/{\it\Gamma}_{0}$.
}
\label{f5-all}
\end{figure}

\begin{figure}[h!]
\centering
\subfigure[\scriptsize{The case $\beta = 1000$, $\eta=0.0001$.}]{
\includegraphics[width=60mm]{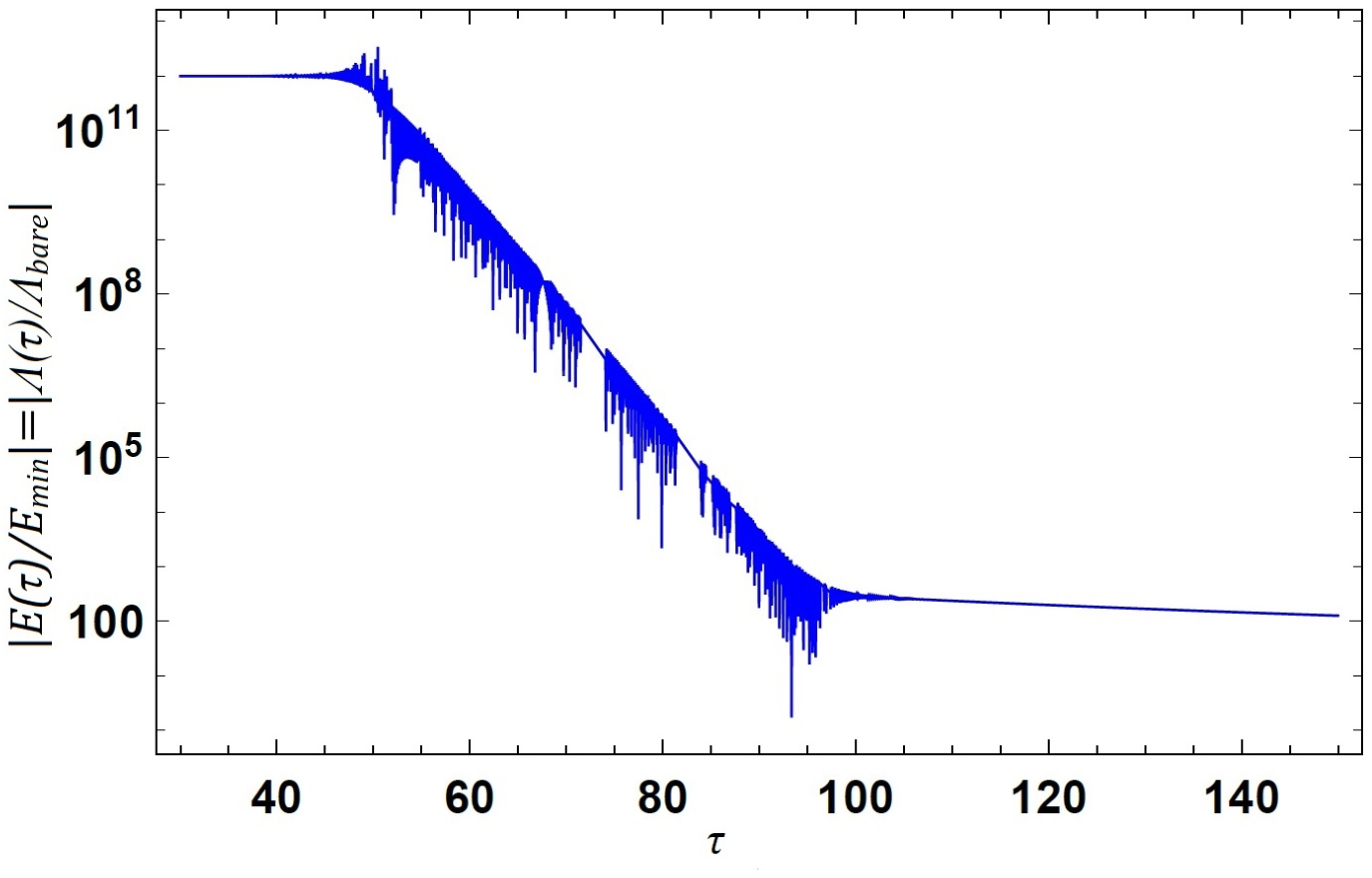}
\label{f61}
}
\;
\subfigure[\scriptsize{The case $\beta= 1000$, $\eta=0.00001$.}]{
\includegraphics[width=60mm]{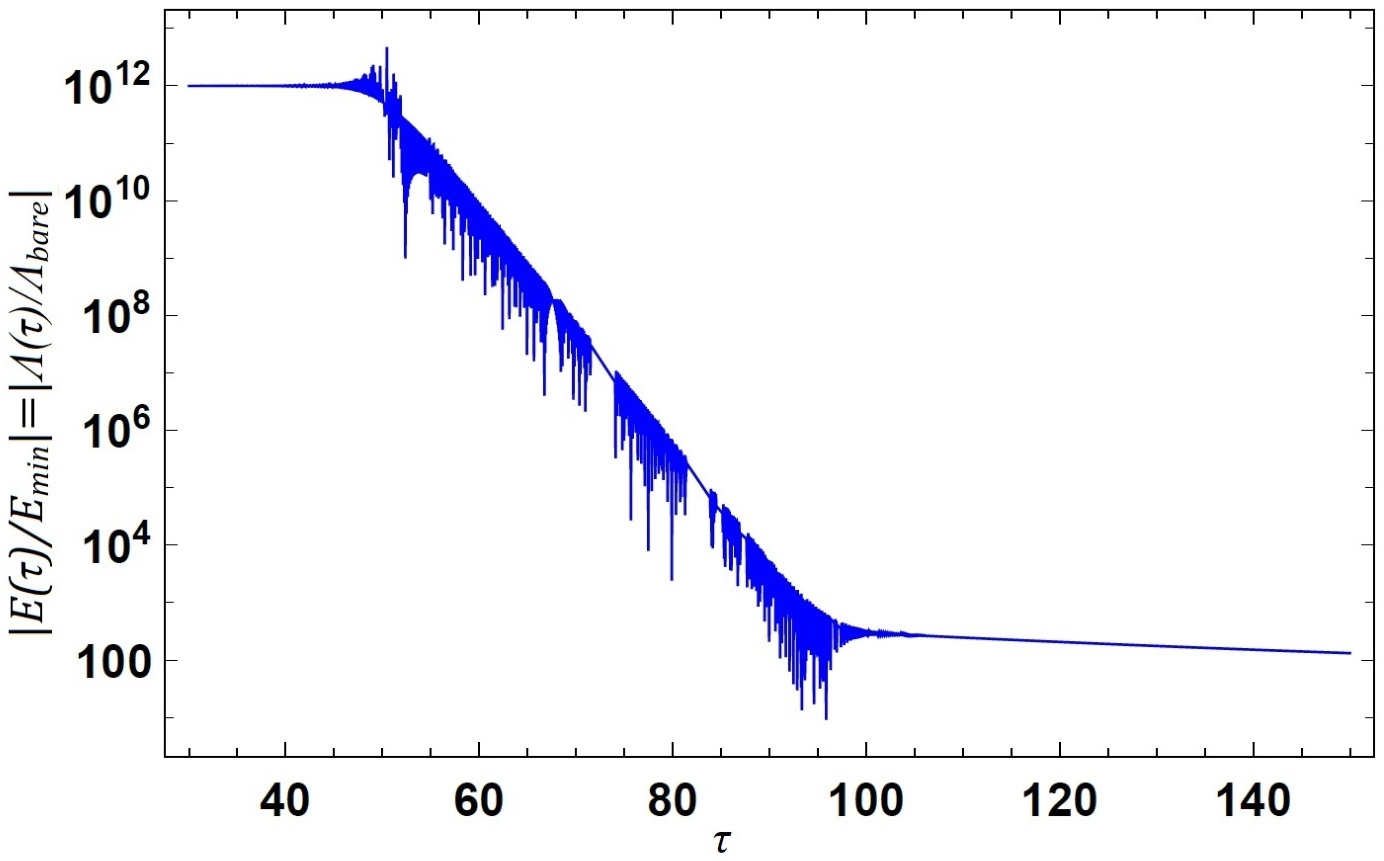}
\label{f62}
}
\caption{\small   A ratio $E(\tau)/E_{min} \equiv \Lambda (t)/\Lambda_{bare}$ obtained for a $\omega(E)$ given by formula (\ref{omega-exp}).
The case $E_{0}/E_{min}=\Lambda_{0}/\Lambda_{bare}= 10^{12}$.
In all figures the time $t$ is measured in lifetimes: $\tau = t/\tau_{0}$: $\tau = t/\tau_{0}$ and $\tau_{0}  = \hbar/{\it\Gamma}_{0}$.
}
\label{f6-all}
\end{figure}

\section{Discussion: possible cosmological applications}

In previous Sections we used the Krylov--Fock theory of unstable states to
find late time properties of the survival amplitude and instantaneous energy. Similar estimations of the late time behavior of the survival amplitude can be also found by means
another method, e.g. methods of the quantum scattering theory. This method was used in \cite{Newton}, where
the long time deviations from the exponential form of the decay law were also studied and  where one can find attempts to estimate the value of $T_{1}$.  Newton on page 624 in Chap.
19 of \cite{Newton} writes:

{\em Let us now take some simple examples. Consider the case of a
nuclear deexcitation by $\gamma$--ray emission. In a typical instance the energy may
be $E_{0} \sim 200 $ keV and the lifetime $\tau  \sim 10^{-8}$ sec so that $\Gamma/E_{0} \sim 3 \times 10^{-13}$.  .... ............}
{\em The decay curve should be
roughly exponential after 1/8 of a mean life from the peak and excellently
exponential after 2. It then remains exponential for $10^{17}$ lifetimes! In order to
destroy most of the exponential--decay curve, one would have to move the
detector away to a distance of about $10^{22}$ miles. ..............}

Note that $10^{22}$ miles   equals $1.7 \times 10^{9}$ light  years approximately, so the effect described by Newton
and in other papers
should be  visible  when analyzing the spectrum of the electromagnetic radiation emitted by cosmic  objects at a distance of $ 1.7 \times 10^{9}$ light  years or more from the Earth's
observer.  Of course in this case it is practically impossible to restore the late time form of the decay curve by means measurements
but as it has already been shown for times $t > T_{1}$ the following quantum effect should take place during  the late time phase
of the quantum decay process:
 The energy of the system in the initial  metastable state, which is approximately equal $E_{0}$ at canonical decay times,
 is forced to decrease as $1/t^{2}$ to
 the minimal energy $E_{min} < E_{0}$.
 This effect was described
 in previous sections and earlier in \cite{Urbanowski:2006mw,Urbanowski:2008kra,Urbanowski:2009lpe,Urbanowski:2011zz,Urbanowski:2016pks,ku-plb-2014a}. In the case of the emission of
 the electromagnetic radiation the excited atomic level can be considered as  an the initial metastable state
 of the system. (In Newton's example the excited state of a nucleus
 is the initial state). So, in the case of very distant cosmic objects emitting electromagnetic radiation
this effect should contribute into the redshift making it apparently larger than it really is.
Such a possibility seems to be very important as this effect may distort the observational results (red shift, luminosity, and so on) and thus lead to wrong conclusions
(For example
estimations of a tension of the Hubble parameter are based on the measurements of the red shift of distant astrophysical objects, etc.).
Possible changes of the red shift caused by this possible effect are described in details in \cite{Urbanowski:2006mw} (see also \cite{ku-plb-2014a}), where
an influence of this property on
measured values of possible deviations of the fine structure constant
$\alpha$ as well as other astrophysical and cosmological parameters were studied
and  this is why this is only signalized here.


Let us analyze now results presented in Fig (\ref{fg3}), Fig (\ref{f1-all}) --- Fig (\ref{f7-all}).
In Figure (\ref{fg3}), in contrast to Figure (\ref{fg2}), it can be seen how quickly in the transition time region, $T_{1} < t < T_{2}$,
 the energy $E (t)$ is reduced
 to its late time  asymptotic form,  $E_{lt}(t) = E_{min} + \alpha_{2}/t^{2} + \ldots$. Namely, if to compare Fig (\ref{fg3}), Fig (\ref{f2-all}) --- Fig (\ref{f7-all})
and Fig (\ref{f1a}) a conclusion can be drawn  that at transition times $T_{1} < t < T_{2}$ the instantaneous energy $E(t)$, (where $E(t) \simeq E_{0}$ for times $t< T_{1}$ and $t
>T_{0}$),  decreases like  an oscillatory modulated exponential function until reaching its asymptotic form $E_{lt}(t)$.
This quantum effect is very strong and efficient: As can be seen from Fig (\ref{fg3}), Fig (\ref{f2-all}) --- Fig (\ref{f7-all}), the reduction of energy $E(t) \simeq E_{0}$ (for $t <
T_{1}$) depending on the values of parameters $\beta$ and $\eta$, may be more than 10 orders: It can be $E(T_{1}) / E_{lt}(T_{2}) \sim 10^{12}$ --- see Fig (\ref{f24}), Fig
(\ref{f71}), and more.
By selecting the appropriate parameters $\beta$  in $\omega_{BW}(E)$, or $\beta$ and $\eta$ in $\omega_{\eta}(E)$, or  the other, more appropriate, energy density distribution
function $\omega(E)$, even much greater  reduction of the energy $E(t) \simeq E_{0}$ can be achieved when time $t$ runs from $t= T_{1}$ to $t=T_{2}$.
Potentially, this effect can reduce the energy by tens of orders or more.
Taking this property into account it seems to be reasonable to hypothesize that this quantum effect can help to explain the cosmological constant problem,
which is consequence of the interpretation
of the dark energy as the vacuum energy: The observed present
value of the cosmological constant is 120 orders of magnitude
smaller than we expect from quantum physics calculations.

One can meet in the large literature cosmological models with metastable vacuum (see, eg. \cite{Rubio:2015zia,Kennedy:1980cj,Branchina:2013jra,Branchina:2014rva} and many others).
Some of these models admit the lifetime of the Universe to be very small \cite{Kennedy:1980cj} or even smaller than the Planck time (see \cite{Branchina:2013jra,Branchina:2014rva}).
 Of course this decaying vacuum is described by the quantum state corresponding to a local minimum of the energy density which is not the absolute minimum of the energy density of the
 system considered.
  In such a case the formalism described in this paper is fully applicable. Let us consider now a cosmological scenario in which
  the lifetime of the false vacuum is shorter than the duration of the inflation phase
   and its decay process began before (or just before) the beginning of the inflation phase and then it is
  continued during the inflationary epoch and later.
This scenario
corresponds with the hypothesis analyzed by Krauss and Dent \cite{Krauss:2007rx,Krauss:2008pt}.
Their hypothesis
suggests that some false vacuum regions do survive well up to the time $T_{1}$ or later.
So, let $|\phi\rangle = | 0\rangle^{\text{F}}$ be a false and $|0\rangle^{\text{T}}$ true vacuum states, respectively, and $E_{0} = E^{\,{\text{F}}}_{0}$ be the energy of a state
corresponding to the false vacuum measured at the canonical decay times, which leads to the vacuum energy density calculated using quantum field theory methods. Let
$E^{\,\text{T}}=E_{min}$ be the energy of true vacuum (i.e., the true ground state of the system).
The fact that the decay of the false vacuum  is the quantum decay process \cite{Coleman:1977py,Callan:1977pt,Krauss:2007rx,Krauss:2008pt,winitzki}
means that state vector corresponding  to the false vacuum is a quantum unstable (or metastable) state. Therefore all the general properties of quantum unstable systems  must also
occur
 in the case of such a quantum unstable state as the false vacuum.
This applies in particular to such properties as late time deviations from the exponential decay law and properties of the energy $E(t) = E^{\,\text{F}}(t)$  of the
system in the quantum false vacuum  state.  In \cite{Urbanowski:2011zz} it was pointed out the energy of those false vacuum regions which survived up to $T_{1}$ and much later differs
from $E^{\,\text{F}}_{0}$.

If one wants to generalize the above results obtained on the basis of quantum mechanics to quantum field theory one should take into account among others a volume factors so that
survival probabilities per unit volume  should be considered and similarly the energies and the decay rate: $E \mapsto  \rho (E) = \frac{E}{V_{0}}$, ${\it\Gamma}_{0} \mapsto \gamma =
\frac{{\it\Gamma}_{0}}{V_{0}}$, where $V_{0} = V(t_{0}^{init})$ is the volume of the considered system at the initial instant $t_{0}^{init}$, when the time evolution starts.
The volume $V_{0}$ is used in these considerations because
the initial unstable state $|\phi\rangle \equiv |0\rangle^{\text{F}}$ at $t=t_{0}^{init}=0$ is expanded into eigenvectors $|E\rangle$ of $\mathfrak{H}$,
 (where $E \in \sigma_{c}(\mathfrak{H})$),  and then this expansion is used to find the density of the energy distribution $\omega (E)$  at this initial instant $t_{0}^{init}$.
Now, if we identify $\rho_{{de}}(t_{0}^{init})$ with the energy $E_{0}^{\;\text{F}}$ of the unstable system divided by the volume $V_{0}$:
$\rho_{{de}}(t_{0}^{init})  \equiv \rho_{0}^{\,\text{F}} \equiv \rho_{0}^{{qft}} \stackrel{\rm def}{=} \rho_{{de}}^{0} = \frac{E_{0}^{\,\text{F}}}{V_{0}}$ and $ \rho_{{bare}}
=\frac{E_{{min}}}{V_{0}}$, (where $\rho_{{0}}^{{qft}}$ is the vacuum energy density calculated using quantum field theory methods) then
it is easy to see that the mentioned changes
$E \mapsto  \frac{E}{V_{0}}$ and  ${\it\Gamma}_{0} \mapsto \frac{{\it\Gamma}_{0}}{V_{0}}$ do not changes the parameter $\beta$:
\begin{equation}
\beta = \frac{E_{0}^{\,\text{F}} - E_{{min}}}{{\it\Gamma}_{0}} \equiv  \frac{\rho_{de}^{0} - \rho_{bare}}{{\gamma}_{0}} > 0, \label{beta-rho}
\end{equation}
(where $\gamma_{0} = {\it\Gamma}_{0}/V_{0}$, or equivalently, $ {\it\Gamma}_{0}/V_{0} \equiv \frac{\rho_{de}^{0} - \rho_{bare}}{\beta}$).
This means that the relations (\ref{E(t)-as}),  (\ref{Re-h-as}), (\ref{kappa}), (\ref{E(t)-Emin}), (\ref{kappa-eta}), (\ref{kappa-eta1}), (\ref{E(t)-Emin-eta}) can be replaced by
corresponding relations for the densities $\rho_{{de}}$ or $\Lambda$ (see, eg., \cite{Urbanowski:2016pks,jcap-2020,Urbanowski:2012pka,ms-ku2}).
Simply, within this approach $E(t)=E^{\,\text{F}}(t)$ corresponds to the running cosmological constant $\Lambda(t)$ and $E_{{min}}$ to the $\Lambda_{bare}$. For example,  we have
\begin{equation}
\begin{split}
\kappa (t) & = \frac{E^{\,\text{F}}(t) - E_{{min}}}{E_{0}^{\,\text{F}} - E_{{min}}}
\equiv \frac{\frac{E^{\,\text{F}}(t)}{V_{0}} - \frac{E_{{min}}}{V_{0}}}{\frac{E_{0}^{\,\text{F}}}{V_{0}} - \frac{E_{{min}}}{V_{0}}} \\
&= \frac{\rho^{F}(t) - \rho_{bare}}{\rho^{F}_{0} - \rho_{{bare}}}
= \frac{\rho_{{de}}(t) - \rho_{{bare}}}{\rho_{{de}}^{0} - \rho_{{bare}}} \\ & = \frac{\Lambda (t) - \Lambda_{{bare}}}{\Lambda_{0} - \Lambda_{{bare}}}.
\label{kappa-L}
\end{split}
\end{equation}
and similarly,
\begin{equation}
\frac{E^{\,\text{F}}(t)}{E_{min}} = \frac{\Lambda (t)}{\Lambda_{bare}},\;\;\;\frac{E_{0}^{\,\text{F}}}{E_{min}} = \frac{\Lambda_{0}}{\Lambda_{bare}}, \label{E-Emin=}
\end{equation}
etc.
Here
$\rho^{F}(t) =  \frac{E^{\,\text{F}}(t)}{V_{0}}\equiv \rho_{{de}}(t) $,  $\Lambda (t) = \frac{8\pi G}{c^{2}}\,\rho_{de}(t)$, (or $\Lambda (t) = 8\pi G\,\rho_{de}(t)$ in $\hbar = c =1$
units), etc.
Equivalently, $\rho_{de}(t) = \frac{c^{2}}{8\pi G} \Lambda(t)$.

Taking into account these relations and analyzing results presented in Fig (\ref{fg3}), Fig (\ref{f1-all}) --- Fig (\ref{f7-all}) one can conclude that within the assumed scenario
there should be,
\begin{equation}
\Lambda (t) \simeq \Lambda_{0} \simeq \frac{8\pi G}{c^{2}}\,\frac{E_{\phi}}{V_{0}} \equiv
\frac{8\pi G}{c^{2}}\,\frac{ ^{\text{F}}\langle 0 |\mathfrak{H}| 0 \rangle^{\text{F}} }{V_{0}}, \;\;\; t\in (T_{0},T_{1}), \label{L-infl}
\end{equation}
at canonical decay times $t < T_{1}$. (Here we used the relation (\ref{D1}) and  the property that $|\Delta_{\phi}^{(1)}| \ll |E_{\phi}|$ from which it follows that in our analysis it
is enough to assume that
$E_{0}^{\,\text{F}} \simeq E_{\phi}$, i.e., that $E_{0}^{\,\text{F}} \simeq\;\,  ^{\text{F}}\langle 0 |\mathfrak{H}| 0 \rangle^{\text{F}}$). In other words there should be $\Lambda
(t) \simeq \Lambda_{0} \equiv \Lambda_{qft} = \frac{8\pi G}{c^{2}}\,\rho_{de}^{qft}$ at times $t < T_{1}$. Then latter, when time $t$ runs from  $t =T_{1}$ to  $t = T_{2}$ the quantum
effect discussed above forces
this $\Lambda(t) \simeq \Lambda_{0}$ to reduce its value  for \linebreak $(t > T_{2})$ to the following one:
\begin{equation}
\Lambda(t) \simeq \Lambda_{eff}(t) = \Lambda_{\text{bare}} + \frac{\alpha_{2}^{\Lambda}}{t^{2}} + \frac{\alpha_{4}^{\Lambda}}{t^{4}} + \ldots \ll \Lambda_{0}.  \label{lambda3}
\end{equation}
Of course in order to reproduce the current value of $\Lambda_{eff}(t)$ from $\Lambda_{qft}$ within the model considered above one should find suitable $\omega (E)$, maybe more
complicated that $\omega_{BW}(E)$ or $\omega_{\eta}(E)$ considered in this paper.
Such a scenario means,  similarly to the idea presented by Krauss and Dent \cite{Krauss:2007rx}
and described  in Sec. 1,
that  the inflation epoch takes during  canonical decay times, $T_{0} < t \leq T_{ 1}$ when
$\rho_{de}(t)\simeq \rho_{de}^{0}$, and thus $\Lambda (t) \simeq \Lambda_{0}$, are extremely large,
then the post--inflationary epoch begins. At the beginning of the
the post--inflationary epoch $\Lambda (t)$ is still very large, but is starting to decrease. It decreases  at times $T_{1} \leq t < T_{2}$
as an oscillatory modulated exponential function to  the value $\Lambda_{eff}(t)$ given by Eq (\ref{lambda3}). Then, at times $t > T_{2}$, $\Lambda (t)$ evolves in  time
as $\Lambda_{eff}(t)$ and tends to $\Lambda_{are}$ as $t \to \infty$.


Einstein's equations
 with the Robertson–-Walker metric
in the standard form of Friedmann equations \cite{Cheng,Sahni}
look as follows:
The first one,
\begin{equation}
\frac{{\dot{a}}^{2}(t)}{a^{2}(t)}  + \frac{kc^{2}}{R_{0}^{2}\,a^{2}(t)} = \frac{8\pi G_{N}}{3}\,\rho +\frac{\Lambda\,c^{2}}{3}, \label{Fr1}
\end{equation}
and the second one,
\begin{equation}
\frac{{\ddot{a}}(t)}{a(t)} =\,-\,\frac{4\pi G_{N}}{3}\,\left(\frac{3p}{c^{2}} +  \rho \right) + \frac{\Lambda\,c^{2}}{3}. \label{Fr2}
\end{equation}
where "dot" denotes the derivative with respect to time $t$, ${\dot{a}}(t) = \frac{d a(t)}{dt}$,
$\rho$ and $p$ are mass (energy) density  and pressure respectively, $k$ denote the curvature signature, and $a(t) = R(t)/R_{0}$ is the scale factor, $R(t)$ is the proper distance at
epoch $t$, $R_{0}= R(t_{0})$ is the distance at the reference time $t_{0}$,
(it
can be also interpreted as the radius of the Universe now)  and here $t_{0}$ denotes  the present epoch. The pressure $p$ and the density $\rho$ are
are related to each other through the equation of state, $p = w\rho \,c^{2}$, where $w$ is constant \cite{Cheng}. There is $w=0$ for a dust, $w=1/3$ for a radiation and $w = -1$ for a
vacuum energy.

Now if the system is in the false vacuum state, $|0\rangle^{F}$,  then at canonical decay times,  $T_{0} < t < T_{1}$, the energy of the system in this state equals
$E^{\,\text{F}}(t) = E_{0}^{\,\text{F}} \approx \,  {^{F}\langle} 0|\mathfrak{H}|0\rangle^{F}$ to a very good approximation and then
 $\Lambda (t) \simeq \Lambda_{0} =  \frac{8\pi G}{c^{2}}\,\frac{^{\text{F}}\langle 0|\mathfrak{H}|0\rangle^{\text{F}}}{V_{0}}> 0$ is very large.
 We assume now that the lifetime of the false vacuum state is much shorter then duration of the inflationary epoch.
 In such a situation the vacuum energy dominates. Therefore in such a case  we can ignore the matter density $\rho$ in Eq (\ref{Fr1}).
In this situation the behavior of expansion rate $\dot{a}(t)$ at times $T_{0} < t < T_{1}$ is such that
the curvature signature  in Eq (\ref{Fr1})  can be always approximated as $ k \approx 0$ (see, e.g. \cite{Cheng}) and then Eq (\ref{Fr1}) simplifies to
\begin{equation}
\frac{{\dot{a}}^{2}(t)}{a^{2}(t)}   \simeq \frac{\Lambda_{0}\,c^{2}}{3}, \;\;\;(T_{0} < t < T_{1}),  \label{Fr1a}
\end{equation}
The solution of this equation  is
\begin{equation}
a(t_{2}) = a(t_{1})\, e^{\textstyle{(t_{2} - t_{1}) \sqrt{\frac{\Lambda_{0}\, c^{2}}{3}}}}, \;\;\;(t_{1},t_{2} \in (T_{0},T_{1})), \label{a-infl}
\end{equation}
which shows that within the considered scenario the scale factor $a(t)$ grows exponentially fast at times, $T_{0} < t < T_{1}$ as it should be at the inflationary epoch. So, in these
times the universe considered, that is the universe born in a metastable false vacuum state  behaves like de Sitter universe.


Note that
if to use the identity $\frac{\ddot{a}(t)}{a(t)} \equiv \frac{\ddot{R}(t)}{R(t)}$ and replace $\frac{\ddot{a}(t)}{a(t)}$
on the left side of equation (\ref{Fr2})  by $\frac{\ddot{R}(t)}{R(t)}$, and then multiply this equation by the product  $m_{p}\, R(t)$, instead of (\ref{Fr2}) we get
an equation that looks like Newton's equation of motion (see, eg. \cite{Cheng,Sahni}),
\begin{equation}
m_{p}\,\ddot{R}(t) = - G\,\frac{m_{p}\;M_{eff}}{R^{2}(t)} + m_{p}\frac{\Lambda\,c^{2}}{3}\,R(t),\label{Fr-N}
\end{equation}
for the point mass $m_{p}$ lying a the sphere with the radius $R(t)$. Here $M_{eff} = \frac{4\pi}{3}\,R^{3}(t)\,(\rho + \frac{3p}{c^{2}})$ is the effective total mass of the sphere of
the radius $R(t) \equiv a(t)\,R_{0}$, $p = w\rho$.  This equation is completely equivalent to equation (\ref{Fr2}). In order to see that it is enough to put $m_{p}=1$ (or to divide it
by $m_{p}$) and then to use $R(t) = a(t)\,R_{0}$. Analyzing
 (\ref{Fr-N})  one can see that the term
$m_{p} \frac{\Lambda\,c^{2}}{3}\,R(t)$ in this  equation plays the same role as a force  in Newton's equations of motion \cite{Sahni}.
And now if there is $\Lambda > 0$  then the force $F =m_{p}\, \frac{\Lambda\,c^{2}}{3}\,R(t) \stackrel{\rm def}{=} F_{rep}$ is a repulsive force and grows with increasing $a(t)$, (or
$R(t) = a(t)\,R_{0}$),  whereas for $\Lambda < 0$  the force $F =\frac{\Lambda\,c^{2}}{3}\,R(t) = F_{att}$ is a force of an attraction \cite{Cheng,Sahni}.

Now basing on these properties of $\Lambda$ we can analyze consequences of the behavior of the energy $E(t) = E^{\,\text{F}}(t)$, and thus $\Lambda (t)$,
at times $T_{1} < t < T_{2}$.
From Figs (\ref{fg2}) and (\ref{f1a}) one can see that for times $T_{1} < t < T_{2}$
there are such time intervals shorter than $(T_{1},T_{2})$, that $E^{\,\text{F}}(t)$ is positive  at some of them and negative for the others. In general $E^{\,\text{F}}(t)$ is
oscillatory modulated at this time region
by changing its value smoothly over time from positive to negative and vice versa.
These properties of  of $E^{\,\text{F}}(t)$  are  reflected in corresponding, analogous behavior of $\Lambda (t)$  on these time intervals: $\Lambda(t)$ is oscillatory modulated for
$t\in (T_{1},T_{2})$.
As a result, acceleration $\ddot{R}(t) \equiv \ddot{a}(t)\,R_{0}$ increases or decreases depending on whether time $t$ runs over the interval with a positive $\Lambda(t)$ or a
negative $\Lambda (t)$ and thus the radius $R(t) = a(t)\,R_{0}$ of the sphere increases slower or faster: Simply our hypothetical sphere  under consideration with the radius $R(t)$ is
vibrating. In other words,  within the considered scenario the Universe in the decaying false vacuum state is pulsating for $t \in (T_{1}, T_{2})$.
This means that the Universe (i.e. the sphere with the radius $R(t) = a(t)\,R_{0}$)  evolving in time and behaving in this way should generate gravitational waves in this phase of its
time evolution when time $t$ runs from $t = T_{1}$  to $t = T_{2}$.
From the point of view of a today's observer,  there is a chance that these relic gravitational waves can be recorded and this is potential observational effect of the scenario
analyzed in this paper.

From Eq (\ref{Fr1}), one more conclusion follows: If to consider the time interval $t \in (T_{1},T_{2})$ only and insert the oscillatory modulated $\Lambda(t)$ into this equation than
one can conclude that the Hubble parameter $H(t) = \frac{{\dot{a}}(t)}{a(t)}$ should also be oscillatory modulated at this time region. So, in general, properties of
$E^{\,\text{F}}(t)$ and thus $\Lambda (t)$ at times $t\in (T_{1},T_{2})$ generated by quantum mechanism considered in this paper correspond with some properties of Early Dark Energy
(EDE) or New Early Dark Energy (NEDE) recently discussed in many papers \cite{VP1,TLS,FN,VP2,EDV,Vag}. The effects predicted by EDE are obtained using the potential $V(\Phi)$  having
an oscillating form, which leads to an energy density $\rho_{de}(t)$ having a similar property at early times.
The advantage of the above--described quantum mechanism over the EDE or NEDE theories lies in the fact that this mechanism requires neither additional fields generating EDE nor
oscillating potentials (see \cite{VP1,TLS,FN,VP2,EDV,Vag}).

Properties of $\Lambda (t)$ for times $t > T_{2}$ are described by Eq (\ref{lambda3}): There is $\Lambda (t) \simeq \Lambda_{eff}(t)$ for $t > T_{2}$, where $\Lambda_{eff}(t)$ is
defined by Eq (\ref{lambda3}). Detailed analysis and discussion of this case can be found in separate papers --- see:
\cite{epjc-2017b,Szydlowski:2017wlv,Szydlowski:2015fya,jcap-2020,ms-ku2}.

\section{Final remarks}

The results and conclusions presented in the previous Section were obtained on the basis of the following assumptions: i)  A transition from the false vacuum state to the true vacuum
state (i.e. the decay of the false vacuum)  is the quantum decay process,  ii) The Universe was born in the false vacuum state, iii) The lifetime of the false vacuum state,  $\tau_{F}
= \frac{ \hbar }{ {\it\Gamma}_{F} }$, (where ${\it\Gamma}_{F}$ is the decay rate, or decay width, of the false vacuum state $|0\rangle^{F}$),  is shorter than the duration of the
inflationary phase of the evolution of the Universe. The picture of the time evolution of the Universe   discussed in Sec. 4 and resulting from these assumptions seems   to be
selfconsistent.

Potentially the effect described and discussed in Sec. 3 and 4 may be considered as a candidate to explain and the problem of the cosmological constant
\cite{Cheng,Weinberg-cos,Weinberg:1988cp,Weinberg:2000yb,Carroll}.
Such a conclusion may be substantiated, for example, by the following analysis: Namely, based
on the  the results presented in Sec. 3
we can estimate the time needed for the value of $\Lambda (t)$ to decrease from the value of $\Lambda_{0}$ to a value close to $\Lambda_{bare}$. So, let us analyze results presented
in Fig  (\ref{f24}). There was assumed that  $\Lambda_{0}/ \Lambda_{bare} = 10^{20}$. From Fig (\ref{f24}) we can conclude that $\Lambda( T_{2})/\Lambda_{bare} \simeq 10^{8}$ and that
$\tau = T_{1}\simeq 50$ and $\tau = T_{2} \simeq 100$. Next using Eq. (\ref{lambda3}) one finds that there is,
\begin{equation}
\frac{\Lambda(\tau)}{\Lambda_{bare}} - 1 = \left(\frac{T_{2}}{\tau}\right)^{2}\,\left[\frac{\Lambda(T_{2})}{\Lambda_{bare}} - 1\right], \label{LL1}
\end{equation}
for $\tau > T_{2}$. Now using the values of $T_{2}$ and of the ratio  $\Lambda( T_{2})/\Lambda_{bare}$ deduced from Fig (\ref{f24}) one concludes that
\begin{equation}
\frac{\Lambda(\tau)}{\Lambda_{bare}} - 1 \approx \left(\frac{ 10^{6}}{\tau}\right)^{2}. \label{LL2}
\end{equation}
This means, e. g. that $\Lambda(\tau)/\Lambda_{bare} \approx 2$ for $\tau \equiv t/\tau_{F} =  10^{6}$. Hence, e.g. if  $\tau_{F} \simeq 10^{-36}$ [s] then
the value \linebreak $\Lambda_{0}/ \Lambda_{bare} = 10^{20}$ taken by $\Lambda (\tau(t))$ for $ \tau < T_{1} \simeq 50$, (in this case $\tau =  T_{1} = 50$ corresponds to $t =
50\tau_{F} =  5\times 10^{-35}$ [s]) reduces to the value $\Lambda(\tau(t))/\Lambda_{bare} \approx 2$ at time $t \approx 10^{6} \tau_{F} =  10^{-30}$ [s]. Analogously, if to assume
that $\tau_{F} \simeq 10^{-38}$ [s] (see, e. g.  \cite{Cheng}, Chap. 9) then the value $\Lambda(\tau(t))/\Lambda_{bare} \approx 2$ can be reached in $t\approx 10^{-32}$ [s].
Analyzing results presented in Figs (\ref{f2-all}) and  (\ref{f2a-all}) one can conclude that the degree of reduction of the energy $E(t)$ (or $\Lambda (t)$)  does not depend on a the
ratio $\Lambda_{0}/\Lambda_{bare}$ but on the magnitude of the coefficient $\beta$. Therefore one can expect that for $\beta = 10^{4}$ (as it is presented in Fig (\ref{f24}) and Fig
(\ref{f2a-all})) there should be $\Lambda (T_{1})/\Lambda (T_{2}) \sim 10^{12}$ for, e.g., $\Lambda_{0}/\Lambda_{bare} \sim 10^{100}$ too and $T_{2} \sim 100 \tau_{F}$. In such a case
there is $\Lambda (T_{2})/\Lambda_{bare} \sim 10^{88}$ and thereafter ratio $\Lambda(\tau(t))/\Lambda_{bare}$ can reach its value $\Lambda(\tau(t))/\Lambda_{bare} \simeq 2$ in time $t
\simeq 10^{8}$ [s] if $\tau_{F} \simeq 10^{-38}$ [s] and in time $t\simeq 10^{10}$ [s] if  $\tau_{F} \simeq 10^{-36}$ [s].
This shows that quantum mechanism discussed in this paper is very effective. So, one can expect that for a suitable $\omega (E)$ the degree of the reduction of $\Lambda (\tau)$ can be
much greater. It seems to be possible that assuming $\tau_{F} \sim 10^{-36}$ [s] (or of a similar  order)
this mechanism would be able to reduce even  the value of $\Lambda_{0}/\Lambda_{bare} \sim 10^{120}$ to the value $\Lambda (\tau)/\Lambda_{bare} \sim 2$ no later than  at time $t
\sim5\times 10^{5}$ years.

Similarly it seems that this quantum effect can also help to explain the   H--tension problem \cite{VP1,TLS,FN,VP2,EDV,EDV2,FN2}.
   The only problem is to find a suitable model with the required lifetime, $\tau_{F}$, of the false vacuum state and thus a suitable energy density distribution $\omega(E)$, what
   requires further studies.
Note that cosmological models with lifetime of the false vacuum state even shorter than the Planck time were  considered in the literature (see e. g.
\cite{Branchina:2013jra,Branchina:2014rva}) but such a lifetime seems to be  too short in order that  canonical decay times could coincide with the inflationary phase.

Some hints concerning  values of the basic parameters of the model we are looking for can be found by analyzing the results obtained numerically and presented in Sec. 3. For example,
let us analyze the results presented in Fig (\ref{f24}). From the Eq. (\ref{beta-rho}) one finds that
$
{\it\Gamma}_{F} = \frac{1}{\beta}(E_{0}^{\,\text{F}} - E_{min}).
$
The results presented in Fig (\ref{f24})  were obtained with the assumption that $\beta = 10^{4}$  and $\frac{E_{0}}{E_{min}} \equiv \frac{E_{0}^{\,\text{F}}}{E_{min}} = 10^{20} $.
Hence $E_{min} = 10^{-20} \,E_{0}^{\,\text{F}} \equiv  10^{-20} \,E_{0}$, and
$
{\it\Gamma}_{F} = {\it\Gamma}_{0} = 10^{-4} \,\times$   $ (1-10^{-20})\,E_{0} \simeq 10^{-4} E_{0} \approx 10^{-4} \langle \phi |\mathfrak{H}|\phi\rangle.
$
Here the approximation $E_{0} \approx  \langle \phi |\mathfrak{H}|\phi\rangle$ was used, which is sufficient
in order to find approximate suitable values of  parameters of the model --- see explanations below Eq. (\ref{L-infl}).
This means that there should be
$E_{min} \approx  10^{-20}\, \langle \phi |\mathfrak{H}|\phi\rangle$,
where $|\phi\rangle$ is the decaying metastable state.
(There is $|\phi\rangle = |0\rangle^{F}$ in the considered case).
Thus, it can be expected that the quantum field theory model with the Hamiltonian $\mathfrak{H}$, in which  the following approximate relations  will take place,
 $E_{0} \approx \langle \phi |\mathfrak{H}|\phi\rangle$, $ {\it\Gamma}_{0} \approx 10^{-4}\, \langle \phi |\mathfrak{H}|\phi\rangle$, and
 $E_{min} \approx  10^{-20}\, \langle \phi |\mathfrak{H}|\phi\rangle$,
  will adequately reflect the process presented in Fig (\ref{f24}).
  There are observational data that
  can be used to constrain possible parameters of the theoretical model that implements the scenario described in this paper, i.e. the form of the potential  $V (\Phi)$ and thus the
  Lagrangian and the Hamiltonian $\mathfrak{H}$, which
 we  are looking for.  Namely, from cosmological observations we have estimations of the time at which inflationary process begins and ends. These times can be used to limit the
 length of the lifetime, $\tau_{F}$, of  the false vacuum: The scenario described above can be realized only if  the lifetime, $\tau_{F}$, is comparable to the time at which the
 inflationary process begins and it is  significantly shorter than the time at which the inflation ends. Having constraints on the lifetime we have a constrain on the decay rate
 ${\it\Gamma}_{F} = {\it\Gamma}_{0} = \hbar/ \tau_{F}$.
  Of course, apart from these conditions, $\omega (E)$ resulting from the properties of such  Hamiltonian $\mathfrak{H}$ must lead to $\Lambda (t)$ satisfying the constraints imposed
  by Eq (\ref{Fr1}). Namely the amplitude $|\Lambda (t)|$ of possible variations of $\Lambda (t)$ at times $t \in (T_{1}, T_{2})$ must be such that the condition $H^{2}(t) > $  is
  satisfied.

One more remark concerning Eq (\ref{Fr1a}),  (\ref{a-infl}) resulting from properties of the energy of very short living metastable false vacuum
and possible connection of them with the inflationary process. From discussion presented in the previous  Section it follows that at times $t < T_{1}$  the energy density $\rho^{F}(t)
\sim \rho_{0}^{F} \simeq \frac{ ^{\text{F}}\langle 0 |\mathfrak{H}| 0 \rangle^{\text{F}} }{V_{0}}$ can be very large. Using equation of state
$p(t) = - \rho^{F}(t)\,c^{2}$ one finds that in this epoch, when $ t < T_{1}$, the pressure $p(t) = p = - \rho_{0}^{F}\,c^{2}$ can take a huge negative values
As a result of which the scale factor $a(t)$ grows exponentially fast
for $t < T_{1}$, which is reflected in the solution (\ref{a-infl}) of equation (\ref{Fr1a}). This effect is exactly what is needed in the inflation process (see, e.g.
{\cite{Cheng,guth}). Next, at times $t \in (T_{1}, T_{2})$ the density $\rho^{F}(t)$ has an oscillatory form and it is still large but decreases  to the values $\rho^{F}(t) -
\rho_{bare} \sim \alpha^{\rho} _{2} /t^{2}$ for times $t > T_{2}$. This means that the scale factor $a(t)$ continues to rise rapidly, but slower and slower when time $t$ runs form $t
= T_{1}$ to $t=T_{2}$ and the time $t = T_{2}$ can be considered as the end of the rapid expansion process.
So, there is potential possibility that this effect  can drive the inflation process.
Concluding:
If the lifetime, $\tau_{F}$, is suitably small then potentially contribution of
this effect  into the inflation process can be significant or even it
 can be responsible for this process.
 The answer to the question whether it is so  and what problems this mechanism solves depends on finding the appropriate potential $V(\Phi)$  and thus the Hamiltonian $\mathfrak{H}$,
 which requires further researches.
Details concerning the  case
 $\rho^{F}(t) - \rho_{bare} \sim \alpha^{\rho} _{2} /t^{2}$ can be found in \cite{epjc-2017b,Szydlowski:2017wlv,Szydlowski:2015fya,jcap-2020}.

Some time one may ask if a transition from the metastable false vacuum state to the true vacuum state, i.e. to the state corresponding with to the absolute minimum of the energy of
the system considered, which is realized as a quantum tunneling process, can be correctly described within the Fock--Krylov theory of quantum decays. The answer is yes, it can be.
In general within the quantum theory  the quantum tunneling is used to model some quantum decay processes, e.g. the process of $\alpha$--decay (see, e.g. \cite{Gamov}), and these
decay processes can be also described using the Fock--Krylov theory (see e.g. \cite{Winter:1961zz,vanDijk:1999zz,vanDijk:2002ru,Dicus:2002tdq,Martorell:2009qpd,vanDijk:2019}).
Strictly speaking in the case of the quantum tunneling used to model the quantum decay process the survival probability can be also expressed in the form of the Fourier transform as
it was done in Eq.~(\ref{a-spec}). This means that the general formalism based on the Fock--Krylov theory is fully suitable for describing the properties of a decaying false vacuum
and a running dark energy. This is why the Fock--Krylov theory was used by Krauss and Dent to analyze late time behavior of false vacuum decay \cite{Krauss:2007rx,Krauss:2008pt},
which papers were an inspiration for our studies.

It should be emphasized here that the conclusions resulting from the formalism used in this paper apply not only to the quantum metastable state of the system prepared at the initial
moment at the local minimum of the potential $V(\Phi)$, but also to other types of metastable false vacuum states, e.g. false vacuum states considered in \cite{guth}.

Cosmological model  with  running $\Lambda$ having the form $\Lambda = \Lambda (t) = \Lambda_{eff}(t)$, where $\Lambda_{eff}(t)$ is given by Eq. (\ref{lambda3}), was discussed in
details in \cite{jcap-2020}, where it was shown that this  $\Lambda_{eff}(t) $ should approximate well $\Lambda (t)$ for times $t > T_{2}$. (In \cite{jcap-2020} the time $T_{2}$ is
denoted as $T_{q-c}$, and, among others, different time scales in the process of decaying metastable dark energy are discussed). It is also shown in \cite{jcap-2020} that
the good approximation of eq.~(\ref{lambda3}) valid for times $t > T_{2}$ is to replace the cosmological time $t$ in it with the Hubble cosmological scale time
$t_\text{H}=\frac{1}{H}$. As the result, instead of (\ref{lambda3}) one gets
\begin{equation}
\Lambda(t)= \Lambda(H(t)) \simeq \Lambda_{\text{bare}} + \alpha_{2} \left(H(t)\right)^{2} +
\alpha_{4} \left(H(t)\right)^{4} + \cdots \, , \label{L(H)}
\end{equation}
that is exactly the parameterization considered in \cite{Shapiro:2003ui,EspanaBonet:2003vk} and in many papers of these and other authors.

Generally, cosmological models with decaying (or running)  $\Lambda = \Lambda (t)$ were considered by many authors (see e.g. \cite{Over} and references therein,
 \cite{maxim,bamba,graef},
 and also \cite{Sahni}) but
the use of the decaying $\Lambda (t)$ by them  was not motivated by the properties of the false vacuum as a quantum unstable state.
The advantage of the method used in this paper on other methods is that we do not assume the form of running $\Lambda (t)$, but   we derive properties of $\Lambda (t)$ and its form,
e.g.  like $\Lambda_{ef}(t)$  for $t > T_{2}$, from the basic assumptions of quantum theory using   the assumption that the transition from a false vacuum to a true vacuum is the
quantum decay process.\\

\noindent
{\bf Acknowledgments:}
This paper is dedicated to the memory of my dear friend and colleague Marek Szyd{\l}o\-wski.

I would like to thank Marek Nowakowski  for  his valuable
comments and discussions.

\hfill\\
{\bf The author contribution statement:} The author declares that there are no conflicts of interest
regarding the publication of this article  and that all results presented in this article are the author's own results.


\begin{thebibliography}{60}

\bibitem{WW}
V. F. Weisskopf, E. T. Wigner,
\emph{Berechnung der nat\"{u}rlichen Linienbreite auf Grund der Diracschen Lichttheorie},  {Z. Phys.} \textbf{63}, 54, (1930);
\emph{\"{U}ber die nat\"{u}rliche Linienbreite
in der Strahlung des harmonischen Oszillators}, Zeit. f. Phys., \textbf{65}, 18, (1930).

\bibitem{Gamov}
G. Gamow,
\emph{Zur Quantentheorie des Atomkernes},  Zeit. f. Phys. \textbf{51}, 204 ---212, (1928).



\bibitem{Coleman:1977py}
S.~R. Coleman, \emph{{The fate of the false vacuum. 1. Semiclassical theory}},
  {Phys. Rev.} {\bfseries D15}  2929, (1977).

\bibitem{Callan:1977pt}
C.~G. Callan, Jr. and S.~R. Coleman, \emph{{The fate of the false vacuum. 2.
  First quantum corrections}},
  {\emph{Phys. Rev.} {\bfseries
  D16},  1762, (1977)}.


\bibitem{rutheford}
E. Ruthheford, \emph{Radioactivity produced in substances by the action of thorium compounds}, {Philosophical Magazine} {\bf XLIX}, 1, 161, (1900).
\bibitem{rutheford1}
E. Rutherford, F. Soddy,\emph{The cause and nature of radioactivity. --Part I}, Philosophical Magazine {\bf  IV}, 370–--396, (1902);
\emph{The cause and nature of radioactivity. --Part II}, Philosophical Magazine {\bf  IV},  569 --- 585,  (1902).
\bibitem{rutheford2}
E. Rutherford, \emph{Radioactive substances and their radiations},  Cambridge Unversity Press 1913. 


\bibitem{Khalfin}
L.~A. Khalfin, \emph{{Contribution to the decay theory of a quasi-stationary
  state}}, {Zh. Exp. Teor. Fiz.} {\bfseries 33} 1371, (1957) .



\bibitem{Fonda}
L.~Fonda, G.~C. Ghirardi and A.~Rimini, \emph{{Decay Theory of Unstable Quantum
  Systems}},
  {Rept.
  Prog. Phys.} {\bfseries 41}  587, (1978).


\bibitem{Peshkin}
M.~Peshkin, A.~Volya and V.~Zelevinsky, \emph{{Non-exponential and oscillatory
  decays in quantum mechanics}},
  {Europhys. Lett.}
  {\bfseries 107}, 40001,  (2014).



\bibitem{misra}
B. Misra and E. C. G.  Sudarshan, {\em The Zeno’s paradox in quantum theory}, J. Math. Phys. {\bf 18}, 756 -- 763 (1977).

\bibitem{bh}
A. Beige and G. C. Hegerfeldt, {\em Projection postulate and atomic quantum Zeno effect}, Phys. Rev. A, {\bf 53}, 53, (1996).


\bibitem{bh1}
A. Beige, G. C. Hegerfeldt and D. G. Sondermann, {\em Atomic Quantum Zeno Effect for Ensembles and
Single Systems}, Foundations of Physics, {\bf 27}, 1671 --- 1688, (1997).

\bibitem{kk}
K. Koshinoa, A.  Shimizuc, {\em  Quantum Zeno effect by general measurements}, Physics Reports, {\bf 412}, 191 -- 275, (2005).

\bibitem{pf}
P. Facchi and S. Pascazio, {\em Quantum Zeno Phenomena:
Pulsed versus Continuous Measurement}, Fortschr. Phys., {\bf 49}, 941 -- 947, (2001).

\bibitem{WMI}
W. M. Itano, {\em et al}, {\em Quantum Zeno effect}, PhysicaL Review A, {\bf 41}, 2295 -- 2300, (1990);
M. C. Fischer, B. Guti\'{e}rrez--Medina, and M. G. Raizen, {\em Observation of the Quantum Zeno and Anti--Zeno Effects in an Unstable System}, Physical Review Letters, {\bf 87},
040402, (2001);
Wenqiang Zheng, D. Z. Xu, Xinhua Peng, Xianyi Zhou, Jiangfeng Du, and C. P. Sun, {\em Experimental demonstration of the quantum Zeno effect in NMR
with entanglement-based measurements}, Physical Review A, {\bf 87}, 032112, (2013);
Erik W. Streed, {\em et al}, {\em Continuous and Pulsed Quantum Zeno Effect}, Physical Review Letters, {\bf 97}, 260402, (2006);
Y. S. Patil, S. Chakram, and M. Vengalattore, {\em Measurement-Induced Localization of an Ultracold Lattice Gas}, Physical Review Letters, {\bf 115}, 140402, (2015);
A. Signoles, {\em et al}, {\em Confined quantum Zeno dynamics of a watched
atomic arrow}, Nature Physics, {\bf 10}, 715 -- 719, (2014).



\bibitem{Urbanowski:2006mw}
K.~Urbanowski, \emph{{A quantum long time energy red shift: A contribution to
  varying alpha theories}},
  {Eur. Phys. J.}
  {\bfseries C58}  151, (2008).

\bibitem{Urbanowski:2008kra}
K.~Urbanowski, \emph{{General properties of the evolution of unstable states at
  long times}},
  {Eur.
  Phys. J.} {\bfseries D54}, 25,  (2009).

\bibitem{Urbanowski:2009lpe}
K.~Urbanowski, \emph{{Long time properties of the evolution of an unstable
  state}},
  {Cent. Eur. J.
  Phys.} {\bfseries 7}, 696,  (2009).

\bibitem{Urbanowski:2011zz}
K.~Urbanowski, \emph{{Comment on `Late time behavior of false vacuum decay:
  Possible implications for cosmology and metastable inflating states'}},
  {Phys. Rev. Lett.}
  {\bfseries 107},  209001, (2011).

\bibitem{Urbanowski:2016pks}
K.~Urbanowski, \emph{{Properties of the false vacuum as the quantum unstable
  state}},
  {{Theor. Math., 458,
  Phys.} {\bfseries 190} (2017).}


\bibitem{Krauss:2007rx}
L.~M. Krauss and J.~Dent, \emph{{The late time behavior of false vacuum decay:
  Possible implications for cosmology and metastable inflating states}},
  {{Phys. Rev. Lett.}
  {\bfseries 100}, 171301,  (2008).}

\bibitem{Krauss:2008pt}
L.~M. Krauss, J.~Dent and G.~D. Starkman, \emph{{Late time decay of the false
  vacuum, measurement, and quantum cosmology}},
  {{Int. J. Mod. Phys.}
  {\bfseries D17},  2501, (2009).}





\bibitem{Cheng}
Ta--Pei Cheng, \emph{Relativity, Gravitation,
and Cosmology:
A basic introduction}, Oxford University Press, 2005.

\bibitem{Weinberg-cos}
S. Weinberg, \emph{Cosmology}, Oxford University Press, 2008.



\bibitem{Landim:2016isc}
R.~G. Landim and E.~Abdalla, \emph{Metastable dark energy},  Phys. Lett.,
  {\bfseries B764},  271 --- 276, (2017).



\bibitem{Stojk}
D. Stojkovic, Glenn D. Starkman, and Reijiro Matsuo, \emph{Dark energy, colored anti–-de Sitter vacuum, and the CERN Large Hadron Collider
phenomenology},
Physical Review, {\bfseries D 77}, 063006, (2008).

\bibitem{Lim}
J.A.S. Lima,
\emph{Extended metastable dark energy},
Physics of the Dark Universe, {\bfseries 30}, 100713, 2020.


\bibitem{epjc-2017b}
A. Stachowski1, M. Szydłowski1, K. Urbanowski,
\emph{Cosmological implications of the transition from the false vacuum
to the true vacuum state},  { Eur. Phys. J.} {\bf  C  77}, 357, (2017).



\bibitem{Szydlowski:2017wlv}
M.~Szydlowski, A.~Stachowski and K.~Urbanowski, \emph{{Quantum mechanical look
  at the radioactive-like decay of metastable dark energy}},
  {{Eur. Phys. J.}
  {\bfseries C77},  902, (2017).}



\bibitem{Szydlowski:2015fya}
M.~Szydlowski and A.~Stachowski, \emph{{Cosmological models with running
  cosmological term and decaying dark matter}},
  {{Phys. Dark Univ.}
  {\bfseries 15}, 96, (2017).}

\bibitem{jcap-2020}
M. Szyd{\l}owski, A. Stachowski
and K.  Urbanowski, \emph{The evolution of the FRW universe with decaying metastable dark energy --- a dynamical system analysis},
{ Journal of Cosmology and Astroparticle Physics},
{\bf 04} 029 (2020).


\bibitem{Krylov:1947tmi}
N.~S. Krylov and V.~A. Fock, \emph{{On two main interpretations of energy-time
  uncertainty}}, {{Zh. Eksp. Teor. Fiz.} {\bfseries 17}, 93,  (1947)}.

\bibitem{Fock:1978fqm}
V.~A. Fock, \emph{{Fundamentals of Quantum Mechanics}}. Mir Publishers, Moscow,
  1978.

\bibitem{Kelkar:2010qn}
N.~G. Kelkar and M.~Nowakowski, \emph{{No classical limit of quantum decay for
  broad states}},
  {J. Phys.}
  {\bfseries A43},  385308,  (2010).



\bibitem{Giacosa:2011xa}
F.~Giacosa, \emph{{Non-exponential decay in quantum field theory and in quantum
  mechanics: the case of two (or more) decay channels}},
  {{Found. Phys.}
  {\bfseries 42},  1262, (2012).}


\bibitem{Giacosa:2018dzm}
F.~Giacosa, \emph{{QFT derivation of the decay law of an unstable particle with
  nonzero momentum}},
  {{Adv.
  High Energy Phys.} {\bfseries 2018},   4672051, (2018).}


\bibitem{Paley}
R. E. A. C. Paley, N. Wiener, \emph{Fourier
transforms in the complex
domain}, American Mathematical Society, New York, 1934.



\bibitem{Peres}
A. Peres,
\emph{Nonexponential decay law}
Ann. Phys. {\bf 129}, 33,  (1980).


\bibitem{Sluis}
K.~M. Sluis and E.~A. Gislason, \emph{{Decay of a quantum-mechanical state
  described by a truncated Lorentzian energy distribution}},
  {{Phys. Rev.} {\bfseries
  A43},  4581, (1991)}.


\bibitem{Goldberger}
M. L. Goldberger, K. M. Watson, {\em Collision Theory}, Willey, New
York 1964.


\bibitem{Arbo}
D. G. Arbo, M. A. Castagnino, F. H. Gaioli and S. Iguri,
\emph{Minimal irreversible quantum mechanics.The decay of unstable states},
 Physica,
{\bf A 227},  469 --- 495, (2000).


\bibitem{Wessner}
J. M. Wessner, D. K. Andreson and R. T. Robiscoe,
\emph{Radiative Decay of the 2P State of Atomic Hydrogen: A Test of the Exponential Decay Law},
Phys. Rev. Lett.
{\bf 29},  1126 --- 1128, (1972).


\bibitem{Norman1}
E. B. Norman, S. B. Gazes, S. C. Crane and D. A. Bennet,
\emph{Tests of the Exponential Decay Law at Short and Long Times}, Phys. Rev.
Lett.,
{\bf 60},  2246 --- 2249, (1988).

\bibitem{Greenland}
P. T. Greenland,
\emph{eeking non-exponential decay},
Nature {\bf 335},  298, (1988).

\bibitem{seke}
J. Seke, W. N. Herfort,
\emph{  Deviations from exponential decay in the case of spontaneous emission from a two--level atom},
 Phys. Rev. {\bf A 38}, 833, (1988).

\bibitem{parrot}
R. E. Parrot, J. Lawrence,
\emph{Persistence of exponential decay for metastable quantum states at long times},
Europhys. Lett. {\bf 57},  632 --- 638, (2002).

\bibitem{lawrence}
J. Lawrence,
\emph{Nonexponential decay at late times and a different Zeno paradox,}
Journ. Opt. B: Quant. Semiclass. Opt. {\bf 4}, No 4,  S446 --- S449, (2002).


\bibitem{joichi}
I. Joichi, Sh. Matsumoto, M. Yoshimura,
\emph{ Time evolution of unstable particle decay seen with finite resolution},
Phys. Rev. {\bf D 58},  045004, (1998).

\bibitem{Nowakowski}
N. G. Kelkar, M. Nowakowski and K. P. Khemchandani,
\emph{Hidden evidence of nonexponential nuclear decay},
Phys. Rev.  {\bf C 70}, 024601, (2004).

\bibitem{Nowakowski2}
M. Nowakowski, N. G. Kelkar,
\emph{Long Tail of Quantum Decay from Scattering Data,}
 AIP Conf. Proc. {\bf 1030},  250 --- 255, (2008); arXiv: 0807.5103.


\bibitem{santra}
R. Santra, J. M. Shainline, Ch. H. Greene, Phys. Rev.
\emph{Siegert pseudostates: Completeness and time evolution}, {\bf A 71}, 032703, (2005).


\bibitem{Winter:1961zz}
R.~G. Winter, \emph{{Evolution of a quasi-stationary state}},
  {{Phys. Rev.} {\bfseries
  123},  1503,  (1961)}.



\bibitem{jiitoh}
T. Jiitoh, S. Matsumoto, J. Sato, Y. Sato, K. Takeda, Phys Rev. {\bf A 71},
012109, (2005).

\bibitem{Rothe}
C.~Rothe, S.~I. Hintschich and A.~P. Monkman, \emph{{Violation of the
  Exponential-Decay Law at Long Times}},
  {{Phys. Rev. Lett.}
  {\bfseries 96},  163601, (2006)}.


\bibitem{Urbanowski:1994epq}
K.~Urbanowski, \emph{{Early-time properties of quantum evolution}},
  {{Phys. Rev.} {\bfseries
  A50},  2847, (1994)}.


\bibitem{Giraldi:2015ldu}
F.~Giraldi, \emph{{Logarithmic decays of unstable states}},
  {{Eur. Phys. J.}
  {\bfseries D69}, 5,  (2015)}.

\bibitem{Giraldi:2016zom}
F.~Giraldi, \emph{{Logarithmic decays of unstable states II}},
  {{Eur. Phys. J.}
  {\bfseries D70},  229, (2016)}.


\bibitem{Breit:1936zzb}
G. Breit and E. Wigner, \emph{Capture of Slow Neutrons},
  {{Phys. Rev.} {\bfseries 49}, 519--531,  (1936)}.


\bibitem{nowakowski}
N. G. Kelkar, M. Nowakowski, \emph{No classical limit of quantum decay for broad states},
{J. Phys. A: Math. Theor.}, \textbf{43}, 385308, (2010).

\bibitem{Kelkar2021}
D. F. Ramírez Jiménez and N. G. Kelkar, \emph{Formal aspects of quantum decay}, Phys. Rev. A, {\bf 104}, 022214, (2021).


\bibitem{R-U}
K.~Raczy\'{n}ska and K.~Urbanowski, \emph{Survival amplitude, instantaneous energy
  and decay rate of an unstable system: Analytical results},
  Acta Phys. Polon.
  {\bf B49}, 1683, (2018).




\bibitem{Newton}
R. G. Newton, {\em Scattering Theory of Waves and Particles},
Springer, New York 1982.


\bibitem{ku-plb-2014a}
K. Urbanowski,  K. Raczy\'{n}ska,
\emph{Possible emission of cosmic X- and ă -rays by unstable particles
at late times}, Physics Letters B, {\bf  731}, 236 --- 241, (2014).



\bibitem{Rubio:2015zia}
J.~Rubio, \emph{{Higgs inflation and vacuum stability}},
  {{J. Phys. Conf.
  Ser.} {\bfseries 631}, 012032,  (2015)}.

\bibitem{Kennedy:1980cj}
A.~Kennedy, G.~Lazarides and Q.~Shafi, \emph{{Decay of the false vacuum in the very early universe}},
 {{Phys. Lett.} {\bfseries B99},  38, (1981)}.

\bibitem{Branchina:2013jra}
V.~Branchina and E.~Messina, \emph{{Stability, Higgs boson mass and new
  physics}},
   {Phys. Rev. Lett.} {\bfseries 111},  241801, (2013).

\bibitem{Branchina:2014rva}
V.~Branchina, E.~Messina and M.~Sher, \emph{{Lifetime of the electroweak vacuum
  and sensitivity to Planck scale physics}},
  {{Phys. Rev.}
  {\bfseries D91}, 013003,  (2015)}.


\bibitem{winitzki}
S. Winitzki,
\emph{Age-dependent decay in the landscape},
{Phys. Rev.} {\bf D 77}, 063508,  (2008).


\bibitem{Urbanowski:2012pka}
K.~Urbanowski, M.~Szydlowski,
\emph{Cosmology with a decaying vacuum},
AIP Conf. Proc. \textbf{1514}, 143,  (2012).

\bibitem{ms-ku2}
M.~Szydlowski, A.~Stachowski, K.~Urbanowski,
\emph{Cosmology with a decaying vacuum energy
parametrization derived from quantum mechani},
Journal of Physics: Conference Series {\bf 626}, 012033,  (2015).


\bibitem{Sahni}
V. Sahni and A. Starobinsky, \emph{The Case for a Positive Cosmological
$\Lambda$--term},  International Journal of Modern
Physics D, {\bf  09}, 373 --- 443, (2000).


\bibitem{VP1}
Vivian Poulin, Tristan L. Smith, Tanvi Karwal, and Marc Kamionkowski,
\emph{Early Dark Energy can Resolve the Hubble Tension},
Physical Review Letters, {\bf 122}, 221301,  (2019).

\bibitem{TLS}
Tristan L. Smith,  Vivian Poulin and Mustafa A. Amin,
\emph{Oscillating scalar fields and the Hubble tension: A resolution
with novel signatures},
Physical Review D, {\bf 101}, 063523, (2020).

\bibitem{FN}
Florian Niedermann and Martin S. Sloth,
\emph{New early dark energy},
Physical Review D, {\bf 103}, L041303, (2021).

\bibitem{VP2}
Vivian Poulin, Tristan L. Smith and Alexa Bartlett,
\emph{Dark Energy at early times and ACT: a larger Hubble constant without late--time
priors},
arXiv: 2109.06229v1 [astro--ph.CO], (2021).

\bibitem{EDV}
Eleonora Di Valentino, Olga Mena,  Supriya Pan,  Luca
Visinelli, Weiqiang Yang, Alessandro Melchiorri,
David F.  Mota, Adam G. Riess and Joseph Silk,
\emph{In the realm of the Hubble tension.a
review of solutions},
Class. Quantum Grav., {\bf 38}, 153001, (2021)


\bibitem{Vag}
Sunny Vagnozzi, {\em Consistency tests of $\Lambda$CDM from the early integrated Sachs--Wolfe effect:
Implications for early-time new physics and the Hubble tension},
Physical Review D,  {\bf 104}, 063524, (2021).




\bibitem{Weinberg:1988cp}
S.~Weinberg, \emph{{The cosmological constant problem}},
  {{Rev. Mod. Phys.}
  {\bfseries 61},  1,  (1989)}.

\bibitem{Weinberg:2000yb}
S.~Weinberg, \emph{{The cosmological constant problems}},  in \emph{{Sources
  and Detection of Dark Matter and Dark Energy in the Universe. Proceedings,
  4th International Symposium, DM 2000, Marina del Rey, USA, February 23-25,
  2000}}, pp.~18--26, 2000,


\bibitem{Carroll}
S. M. Carroll,
\emph{The Cosmological Constant}, Living Rev. Relativity,{\bf 4}, 1,  (2001); arrXiv:astro-ph/0004075v2.


\bibitem{EDV2}
E.~Di~Valentino, E.~V. Linder and A.~Melchiorri, \emph{Vacuum phase transition
  solves the $H_0$ tension},
  Phys. Rev.,
  {\bfseries D97}, 043528, (2018).



\bibitem{FN2}
Florian Niedermann and Martin S. Sloth,
\emph{Resolving the Hubble tension with new early dark energy},
Physical Review D, {\bf  102}, 063527, (2020).



\bibitem{vanDijk:1999zz}
W.~van Dijk and Y.~Nogami, \emph{{Novel expression for the wave function of a
  decaying quantum system}},
  {{Phys. Rev. Lett.}
  {\bfseries 83}, 2867,  (1999)}.

\bibitem{vanDijk:2002ru}
W.~van Dijk and Y.~Nogami, \emph{{Analytical approach to the wave function of a
  decaying quantum system}},
  {{Phys. Rev.} {\bfseries C65}, 024608,  (2002)}.

\bibitem{Dicus:2002tdq}
D.~A. Dicus, W.~W. Rebko, R.~F. Schwitters and T.~M. Tinsley, \emph{{Time
  development of a quasistationary state}},
  {{Phys. Rev.}
  {\bfseries A65}, 032116,  (2002)}.

\bibitem{Martorell:2009qpd}
J.~Martorell, J.~G. Muga and D.~W.~L. Sprung, \emph{Quantum post-exponential
  decay},  in \emph{Time in Quantum Mechanics, Vol. 2}, G.~Muga, A.~Ruschhaupt
  and A.~del Campo, eds., vol.~789 of \emph{Lect. Notes Phys.}, (Berlin),
  pp.~239--275, Springer-Verlag, (2009),

\bibitem{vanDijk:2019}
W.~van Dijk and F.~M. Toyama, \emph{{Decay of a quasistable quantum system and
  quantum backflow}},
  {{Phys. Rev.}
  {\bfseries A100},  052101,  (2019)}.


\bibitem{Shapiro:2003ui}
I.~L. Shapiro, J.~Sola, C.~Espana-Bonet and P.~Ruiz-Lapuente, \emph{{Variable
  cosmological constant as a Planck scale effect}},
  {{Phys. Lett.}
  {\bfseries B574}, 149,  (2003)}.

\bibitem{EspanaBonet:2003vk}
C.~Espana-Bonet, P.~Ruiz-Lapuente, I.~L. Shapiro and J.~Sola, \emph{{Testing
  the running of the cosmological constant with type Ia supernovae at high z}},
  {{JCAP} {\bfseries
  0402}, 006,  (2004)}.



\bibitem{guth}
A. H. Guth, \emph{Inflation}, Proc. Nat. Acad. Sci. USA, {\bfseries 90}, 4871 --- 4877, (1993); {\emph{Eternal inflation and its implications}}, J. Phys. A: Math. Theor. {\bfseries
40}, 6811 --- 6826, (2007).


\bibitem{Over}
J. M. Overduin and F. I. Cooperstock, \emph{Evolution of the scale factor with a variable cosmological term},
Physical Review D, {\bf 58}, 043506, (1998).

\bibitem{maxim}
I. G. Dymnikova, M. Yu. Khlopov,
\emph{Self-consistent initial conditions in inflationary cosmology}, Gravitation \&
Cosmology, {\bf 4}, Supplement;   50 --- 55, (1988).
  I. G. Dymnikova, M. Yu. Khlopov,
\emph{Decay of cosmological constant as Bose condensate evaporation}, Mod. Phys.
Lett. A . (2000) {\bf 15}, 2305 --- 2314, (2000); arXiv: astro-ph/0102094.
I. G. Dymnikova, M. Yu. Khlopov,
\emph{Decay of cosmological constant in self-consistent inflation} Eur. Phys. J. C
{\bf 20}, 139 --- 146, (2001).
S. Ray, M. Yu. Khlopov, P. P. Ghosh and Utpal Mukhopadhyay,
\emph{Phenomenology of $\Lambda$-CDM model: a possibility of accelerating
Universe with positive pressure}, Int. J. Theor. Phys. (2011), {\bf 50}, 939 --- 951, (2011); arXiv: 0711.0686 [gr-qc].



\bibitem{bamba}
K. Bamba, S. Capozziello, S. Nojiri and S. D. Odintsov, \emph{Dark energy cosmology: the equivalent description via different
theoretical models and cosmography tests,}
Astrophys. Space Sci.  {\bf 342}, 155, (2012);
[arXiv:1205.3421 [gr-qc]].


\bibitem{graef}
Elcio Abdalla, L. L. Graef, Bin Wang, \emph{A model for dark energy decay}, Physics Letters {\bf B726}, 786 --- 790, (2013).





\end{thebibliography}
\end{document}